\documentclass[aps,prl,10pt,longbibliography,twocolumn,showpacs,amsmath,amssymb,superscriptaddress,floatfix,nobibnotes]{revtex4-2}
\usepackage[utf8]{inputenc}
\usepackage{graphicx}  
\usepackage{soul}
\usepackage{bm}       
\usepackage{amsfonts,amsmath,amssymb,mathtools}  
\usepackage[colorlinks=true,breaklinks=true,allcolors=blue]{hyperref}
\usepackage{orcidlink}
\usepackage{physics}
\usepackage{dsfont}
\usepackage{enumitem}
\definecolor{SR}{rgb}{0.01, 0.75, 0.24}

\begin{document}
\title{Time complexity of a monitored quantum search with resetting}

\author{Emma C. King \orcidlink{0000-0002-6696-3235}}
\thanks{These authors contributed equally to this work.}
\affiliation{Theoretical Physics,  Saarland University,  D-66123  Saarbr\"ucken,  Germany}

\author{Sayan Roy 
\orcidlink{0009-0005-5805-3791}}
\thanks{These authors contributed equally to this work.}
\affiliation{Theoretical Physics,  Saarland University,  D-66123  Saarbr\"ucken,  Germany}

\author{Francesco Mattiotti \orcidlink{0000-0002-2532-8876}}
\affiliation{Theoretical Physics,  Saarland University,  D-66123  Saarbr\"ucken,  Germany}

\author{Maximilian Kiefer-Emmanouilidis \orcidlink{0009-0005-7065-3511}}
\affiliation{Embedded Intelligence, German Research Center for Artificial Intelligence (DFKI), D-67663 Kaiserslautern, Germany}
\affiliation{Department of Computer Science and Research Initiative QC-AI, RPTU Kaiserslautern-Landau, D-67663 Kaiserslautern, Germany }
 
\author{Markus Bl\"aser \orcidlink{0000-0002-1750-9036}}
\affiliation{Computer Science, Saarland University, D-66123 Saarbr\"ucken, Germany}
\affiliation{Center for Quantum Technologies (QuTe), Saarland University, Campus, 66123 Saarbr\"ucken, Germany}

\author{Giovanna Morigi \orcidlink{0000-0002-1946-3684}}
\affiliation{Theoretical Physics,  Saarland University,  D-66123  Saarbr\"ucken,  Germany}
\affiliation{Center for Quantum Technologies (QuTe), Saarland University, Campus, 66123 Saarbr\"ucken, Germany}

\date{\today}

\begin{abstract}  

Searching a database is a central task in computer science and is paradigmatic of transport and optimization problems in physics. For an unstructured search, Grover's algorithm predicts a quadratic speedup, with the search time $\tau(N)=\Theta(\sqrt{N})$ and $N$ the database size. Numerical studies suggest that the time complexity can change in the presence of feedback, injecting information during the search. Here, we determine the time complexity of the quantum analog of a randomized algorithm, which implements feedback in a simple form. The search is a continuous-time quantum walk on a complete graph, where the target is continuously monitored by a detector. Additionally, the quantum state is reset if the detector does not click within a specified time interval. This yields a non-unitary, non-Markovian dynamics. We optimize the search time as a function of the hopping amplitude, detection rate, and resetting rate, and identify the conditions under which time complexity could outperform Grover's scaling. The overall search time does not violate Grover’s optimality bound when including the time budget of the physical implementation of the measurement. For databases of finite sizes monitoring can warrant rapid convergence and provides a promising avenue for fault-tolerant quantum searches.
\end{abstract}

\maketitle

\emph{Introduction.\textemdash}The Grover search algorithm is paradigmatic of quantum supremacy: in an unstructured database of $N$ elements it promises a quadratic speedup, with time complexity scaling as $\tau_G(N)=\Theta(\sqrt{N})$ with respect to the classical search where $\tau_C(N)=\Theta(N)$ \cite{Grover97,Nielsen_Chuang10}. The dynamics realizes amplitude amplification and, in the ideal implementation, the search fidelity periodically reaches unity with the period $\tau_G(N)$. 
The error performed by not stopping the search at the exact time can be bound \cite{Yoder_etal14}.

The digital implementation in terms of a quantum circuit finds an analog counterpart in adiabatic quantum searches \cite{Farhi_etal00,Albash_2018}. The latter can approach asymptotically unit fidelity with timescales $\tau_G(N)$ for appropriate choices of the drive \cite{Roland_Cerf02, Wong_Meyer16}. Adiabatic quantum searches are paradigmatic of quantum annealing \cite{Santoro_2002,Albash_2018}, are relatively robust against non-unitary contributions emerging from stochasticity \cite{Amin_etal08, deVega_etal10, Wild_etal16, Albash_2018} and can profit from non-unitary dynamics \cite{Avron_etal11,Pick:2019, Menu_etal22,Lewalle:2023,Berwald:2024,King:2024,Sveistrys_2025,Lewalle:2024}.

A quantum walk on a graph is a further prominent analog implementation  \cite{Farhi_Gutmann98,Aaronson:2005,Kendon2010}. The mapping permits to interpret the search in terms of physical diffusive processes. In the continuous-time realization, the vertices are the elements of the database and the diffusive dynamics are equivalent to a spatial search \cite{Ambainis:2003,Childs_2004}. As such, they provide important insights into optimization problems, including quantum annealing away from the adiabatic limit \cite{Ambainis:2003, Callison_etal19, Roget_etal20}. As in Grover's digital implementation, the walker's wavepacket periodically alternates localization at the target site with spreading across the full graph. The time complexity of the search achieves Grover's quadratic scaling under optimal conditions \cite{Childs_etal03, Meyer_Wong15, Chakraborty_etal20_2, Lewis_etal21, King:2025}. The dynamics is relatively robust against incoherent contributions \cite{Kendon_Tregenna03,Kendon_Tregenna04} and can be stabilized by protocols based on the quantum Zeno mechanism \cite{Pyshkin:2022,Berwald:2025}. Recent works provided numerical evidence that monitoring and feedback can speed up the convergence of the search over small databases \cite{Candeloro_etal23,Schulz_etal2024_GuidedQuantumWalk,Nzongani:2025}. Moreover, the non-unitary dynamics ideally leads to the asymptotic occupation of the target site, thus lifting the requirement of stopping the search at the exact time. This points toward exploiting non-unitary dynamics for quantum supremacy. This hypothesis is corroborated by recent studies on the scaling with $N$ of the speed of non-unitary protocols for quantum state preparation  \cite{Morigi:2015,Abbasi:2022,
Allahverdyan_Petrosyan2022_DissipativeSearchUnstructured,Ding_etal2025_EndtoEndEfficientQuantum,Zhan_etal2025_RapidQuantumGround,Cubitt2023_DissipativeGroundState, Eder2025, Ding_2024_single_ancilla} and brings to the fore the need to determine the time complexity of feedback-based search protocols, namely, the scaling with $N$ of the non-unitary search dynamics.

\begin{figure*}[t]
    \centering
    \includegraphics[width=\linewidth]{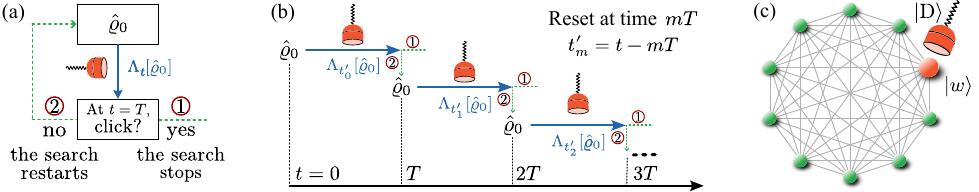}
    \caption{(a) Search dynamics with resetting. A feedback mechanism reinitializes the state to $\hat\varrho_0$ if the detector monitoring the target has not clicked in the time interval $T$.~(b) Dynamics: For $t\in (mT, (m+1)T)$, the dynamics is governed by the CPTP map $\Lambda_{t_m'}[\hat\varrho_0]$ with $t_m'=t-mT$. At times $t= mT$ the protocol either terminates upon a click or otherwise restarts from state $\hat{\varrho}_0$. (c) The search is a continuous-time quantum walk on a globally connected graph with a detector continuously monitoring the target $\vert w\rangle$, corresponding to the orange-colored vertex. }
    \label{fig:1}
\end{figure*}

In this work, we determine the cost function and the time complexity of a relatively simple realization of a non-unitary quantum search with feedback. The protocol is inspired by quantum resetting dynamics \cite{Varbanov:2008,Thiel:2018,Didi2022Measurement-inducedWalks,Friedman2017QuantumProblem,Yin:2025,Dhar2015QuantumModel,Dhar2015,Das2022QuantumChain,Dattagupta:2022, Mukherjee2018QuantumReset} and combines repeated measurement at the target site with resetting to the initial state conditioned on no detection, see Fig.\ \ref{fig:1}(a)-(b). The dynamics is non-unitary and provably non-Markovian \cite{Perfetto2021DesigningResetting,Perfetto2022ThermodynamicsResetting}. The protocol realizes a quantum analog of randomized algorithms \cite{Lu_etal16} and shares key features with Sch\"oning's algorithm~\cite{uwe_satisfiability_2013, eshaghian2024runtimecoherencetradeoffshybridsatsolvers}, which is an effective randomized solver of Boolean satisfiability problems. We note that assessing the computational time complexity of a non-unitary protocol is currently a challenge due to the lack of an established framework for optimizing open quantum system dynamics. Optimization procedures have been implemented for Zeno-based protocols \cite{Avron_2010,King:2024,Lewalle:2023,Lewalle:2024}. Time complexity of Markovian, non-Hermitian dynamics has been discussed in \cite{Barch_Lidar2025_ComputationalComplexityNonHermitian,Childs_2017}. Our protocol is amenable to an analytic solution: this permits us to identify the cost function and to perform optimization of the non-unitary dynamics.

\emph{Monitored search with resetting.\textemdash}The search dynamics is performed on a graph, whose vertices are the elements of the database, represented by $N$ bit strings. The graph is illustrated in Fig.~\ref{fig:1}(c) and is assumed to be complete. The target state $\ket{w}$ is tagged by a different energy and is monitored by a detector. The resetting procedure consists of reinitializing the walker state if the detector has not clicked after a time $T$ has elapsed, see Fig.~\ref{fig:1}(a)-(b). We model the detector by a two-level system whose states $|0_\mathrm{D}\rangle$ and $|1_\mathrm{D}\rangle$ correspond to no click and a click, respectively. The initial state of graph and detector is $\hat\varrho_0=\hat{\rho}_0\otimes|0_\mathrm{D}\rangle\langle 0|$. Here,
$\hat{\rho}_0=|s\rangle\langle s|$ is the uniform superposition of all the graph's vertices, being $\lvert s\rangle = \sum_{j=1}^{N} \lvert j \rangle/\sqrt{N}$, and also the state in which the resetting procedure reinitializes the walker. Resetting occurs at the stroboscopic times  $t_m=mT$ (with $m\in \mathbb{N}$), while the search dynamics between the resetting events $t\in (t_m,t_{m+1})$ is described by a completely positive, trace-preserving (CPTP) map $\Lambda_t$, which acts on the density operator of system and detector: $\hat{\varrho}_{t-t_m} = \Lambda_{t-t_m}[\hat{\varrho}_0]$~(see Fig.~\ref{fig:1}(b)). The map's generator is the Liouvillian 
\begin{eqnarray}
\label{eq:Liouville}
     \mathcal{L}\hat{\varrho}_t &=& -{\rm i} (\hat{H}_{\rm eff} \hat{\varrho}_t-\hat{\varrho}_t \hat{H}_{\rm eff}^\dagger) \\
     & &+ 2 \kappa \vert w, 1_\mathrm{D}\rangle \langle w, 0_\mathrm{D} \vert \hat{\varrho}_t \vert w,0_\mathrm{D} \rangle\langle w, 1_\mathrm{D}\vert\,,\nonumber
\end{eqnarray}
where $\kappa > 0$ is the monitoring rate, such that $\Lambda_t[\hat{\varrho}_0]=\exp(\mathcal L t)\hat{\varrho}_0$. The operator $\hat{H}_{\rm eff}=\hat{H}_{\rm eff}^{(s)} \otimes |0_\mathrm{D}\rangle\langle 0_\mathrm{D}|$ is a non-Hermitian Hamiltonian. The operator defined on the Hilbert space of the graph reads:
\begin{equation}\label{eq:H_complete}
	\hat{H}_{\rm eff}^{(s)} =  - \gamma N|s\rangle\langle s|- (\epsilon_w+i \kappa) \ket{w} \bra{w}\,,
\end{equation}
with the tunneling amplitude $\gamma>0$ and the energy of the target state $\epsilon_w$, which we set $\epsilon_w=1$. The dynamics is in dimensionless units, corresponding to rescaling the time by $\hbar /\epsilon_w$. 

In the absence of monitoring ($\kappa = 0$), the quantum walk's dynamics is unitary and the search is optimal for $\gamma=1/N$ with search time $\tau_G(N)$ \cite{Farhi_Gutmann98,Childs_2004,King:2025}. In the presence of monitoring and resetting, the dynamics at $t\in (t_m,t_{m+1})$ is governed by the map: 
\begin{equation}
    \hat{\varrho}_t = (1-P(t_m)) |w\rangle\langle w|\otimes\vert 1_\mathrm{D} \rangle \langle 1_\mathrm{D}\vert + P(t_m) \Lambda_{t-t_m}[\hat{\varrho}_0]\,,
\end{equation}
where $P(t)={\rm Tr}\{\hat{\varrho}_{t}|0_\mathrm{D}\rangle\langle 0_\mathrm{D}|\}$ is the probability that the detector has not clicked at time $t$ (no-click probability).
The resulting dynamics is a renewal process and is non-Markovian, since it depends on the memory of the last reset event \cite{Perfetto2022ThermodynamicsResetting}. The detection process, in turn, is described by an irreversible decay from the state $|w, 0_\mathrm{D}\rangle$ to the state $|w, 1_\mathrm{D}\rangle$, at which the search naturally stops.

The no-click probability $P(t)$ is central to the dynamics of quantum resetting~\cite{Yin2023RestartTimes, roy2025causalitylocalisationuniversalitymonitored}. It determines the waiting time statistics of the detection events \cite{Cohen-Tannoudji:1986,Dum1992} and therefore the timescale of the search in our system.  It satisfies the recursive relation $P(t)=P(t_m)P(t-t_m)$, which holds for $t\in (t_m,t_{m+1})$. Using that $P(0)=1$, we find $P(t_m)=P(T)^m$, where  $P(t) =  \textrm{Tr}\{|0_\mathrm{D}\rangle\langle 0_\mathrm{D}| {\rm e}^{{\mathcal L}t} \hat{\varrho}_0\}$. Inserting the explicit form of the initial state leads to the compact expression
\begin{equation}
    \label{eq:surv_prob_trace}
    P(t) =  \langle s|\hat S(t)^\dagger\hat S(t)|s\rangle\,,
\end{equation} 
with $\hat S(t)={\rm e}^{-{\rm i}\hat{H}_{\rm eff}^{(s)}t}$. 

\emph{Cost function.\textemdash}The search fidelity is the probability that the walker has reached the target irrespective of the state of the detector, $\mathcal F(t)={\rm Tr}\{|w\rangle\langle w|\hat\varrho_t\}$.  It can be cast into the sum of two components, $\mathcal F(t)=\mathcal F_0(t)+\mathcal F_1(t)$, which measure either the probability that the walker has reached the target conditioned to no detection ($\mathcal F_0=\langle w, 0_\mathrm{D}|\hat{\varrho}_t|w,0_\mathrm{D}\rangle$) or to a detector click ($\mathcal F_1=\langle w, 1_\mathrm{D}|\hat{\varrho}_t|w,1_\mathrm{D}\rangle$). In this second case, the search stops, while in the first case, after reaching the target site, the walker wavepacket might spread again across the graph. Between two resetting events, at time $t\in (t_m,t_{m+1})$, the two components of $\mathcal F(t)$ can be expressed as
\begin{eqnarray}\label{eq:F0_main_text}
	\mathcal F_0(t)&=&P(T)^m|\langle w|\hat{S}(t-t_m)|s\rangle|^2,\\
    \mathcal F_1(t)&=&1-P(T)^m P(t - t_m)\,,\label{eq:F1_main_text}
\end{eqnarray}
and are explicit functions of the no-click probability $P(t)$. Note that $|\langle w|\hat{S}(t)|s\rangle|^2\le P(t)$, therefore $\mathcal F_0(t_m)\le P(T)^m$. Therefore, the optimal strategy consists in optimizing either $\mathcal F_0(t)$ or $\mathcal F_1(t)$, but not both simultaneously. In particular, maximizing $\mathcal{F}_1$ requires minimizing $P(T)$. Within this strategy, the search asymptotically converges to the target.

\emph{Time complexity of monitored search.\textemdash}We define the search time $\tau(N)$ by the minimal time at which the fidelity $\mathcal F(t)$ reaches the threshold value $\mathcal F_{\rm th}=1-\varepsilon$, with $\varepsilon \in[0,1)$. Its scaling with $N$ is the time complexity of the algorithm. In general, the ratio $\gamma/\kappa$ determines whether the dynamics is dominated by tunneling or classical effects \cite{Nzongani:2025}. It is instructive to discuss two limiting cases. For $\kappa=0$ the dynamics is unitary, the no-click probability $P(t)=1$, and $\mathcal F(t)=\mathcal F_0(t)$ is the fidelity of the spatial search of Refs.\ \cite{Farhi_Gutmann98, Childs_2004}.  For $\gamma=1/N$ and $N\gg 1$ the fidelity oscillates with time as $\mathcal F(t)\simeq\sin^2(t/\sqrt{N})$ when the walker is initialized in the state $ |s\rangle$. The search time is $\tau(N)=\pi\sqrt{N}/2$ and the condition $\mathcal F(t)\ge \mathcal F_{\rm th}$ is reached for $|t-\tau|\le \delta \tau$, where the tolerance scales as $\delta \tau\sim 2\sqrt{2\varepsilon}/\pi$ for $\varepsilon \ll N$. In the opposite case, for $\gamma=0$ and $\kappa>0$, the fidelity $\mathcal F(t)=\mathcal F_1(t)$. For $\kappa T\ge 1$ the detection probability is the overlap $1/N$ between the initial state and the target state, and the no-click probability is $P(T)=1-1/N$. Convergence to the target is guaranteed by alternating initialization and resetting, analogous to an optical pumping mechanism \cite{Kastler:1957}.  The fidelity $\mathcal F(t)$ is maximal after $m_0$ resetting events, when $P(T)^{m_0}=\varepsilon$. For $N\gg 1$ then $\tau=m_0T\approx N|\ln \varepsilon| T$. For a fixed resetting time $T$, the time complexity is the linear scaling $\Theta(N)$ of a classical search. 

We analyze the time complexity of the monitored search by maximizing the fidelity ${\mathcal F}_1$ for $\gamma,\kappa>0$. The functional behavior of the probability $P(T)$ determines the time dependence of ${\mathcal F}_1$ and is obtained by diagonalizing the non-Hermitian Hamiltonian $\hat{H}_{\rm eff}^{(s)}$ over the two-dimensional Hilbert space spanned by the orthonormal basis $\{|w\rangle,|r_\perp\rangle\}$ with $|r_\perp\rangle=\sum_{j\neq w}|j\rangle/\sqrt{N-1}$. The eigenvalues read 
\begin{equation}
    \lambda_\pm = \left(a+d\pm
	\sqrt{(a-d)^2+4b^2}\right)/2\,,
\end{equation}
with $a \equiv -(\gamma+1+i\kappa)$, $b \equiv -\gamma\sqrt{N-1}$ and $d\equiv -\gamma(N-1)$, and are degenerate at the exceptional point $\gamma^{\rm EP}=1/(N-2)$ and $\kappa^{\rm EP}=2\gamma^{\rm EP}\sqrt{N-1}$. The corresponding right and left eigenvectors respectively satisfy the relations $\hat{H}_{\rm eff}^{(s)}|\lambda_{\pm}\rangle=\lambda_{\pm}|\lambda_{\pm}\rangle$ and $\langle\bar\lambda_{\pm}|\hat{H}_{\rm eff}^{(s)}=\lambda_{\pm}\langle\bar\lambda_{\pm}|$~\cite{morse_methods_1953, brody_biorthogonal_2014}. Moreover, $\langle\bar\lambda_\pm|=|\lambda_\pm\rangle^{T}$ due to the symmetry of the Hamiltonian, $\hat{H}_{\rm eff}^{(s)}=(\hat{H}_{\rm eff}^{(s)})^{T}$. Away from the exceptional point they form a biorthogonal basis and the no-click probability takes the compact form
\begin{equation}\label{eq:surv_prob_main}
    P(t) = \sum_{j,k=\pm}e^{-{\rm i}(\lambda_j - \lambda_k^*)t}b_{jk} \langle s \vert \bar{\lambda}_j \rangle \langle \bar{\lambda}_k |s\rangle
\end{equation}
with $b_{jk}=\langle \lambda_j \vert \lambda_k \rangle$. The diagonal terms $j=k$ in~\eqref{eq:surv_prob_main} decay as $e^{-2|\mathrm{Im}\lambda_j| t}$ with rate $2|\mathrm{Im}\,\lambda_j|$ while the off-diagonal terms describe damped oscillations, see the Supplemental Material (SM) \cite{SM2}. The problem is equivalent to a two-level transition between the state $|r_\perp\rangle$ and an unstable state $|w\rangle$ with decay rate $\kappa$. The amplitude of the coherent coupling is $\gamma\sqrt{N-1}$ and the energy offset (gap) is $\Delta=|\gamma (N-2)-1|$.  The scaling with $N$ of $\gamma$ and $\kappa$ is key to determining the time complexity of the dynamics. 
\begin{table}[t]
    \caption{Scaling exponent $\alpha$ of the time complexity, $\tau = \Theta(N^{\alpha})$, for different parameter regimes of the parametrization \eqref{eq:scaling_rs}. Note that the time complexity at $\bar r=0$, regimes (A1) and (A2), depends on the value of $\bar\gamma$; see SM \cite{SM2, SM4}.}
    \centering
    \setlength{\tabcolsep}{4pt}
    \begin{tabular}{c|c|c|c|c|c}
    \hline\hline
     Regime &$\gamma \sim N^{-(\bar{r} + 1)}$&$\bar{\gamma}$&$\kappa \sim N^{-s}$&$\bar{\kappa}$&$\mathbf{\alpha}$\\
      \hline
      (A1)& $\bar{r}= 0$ &$1$& $s\geq1/2$&$\mathcal{O}(1)$&$s$\\
      (A2)& $\bar{r}= 0$&$1$&$s\leq1/2$&$\mathcal{O}(1)$&$1-s$\\
      (B)& $\bar{r}>0$&$\mathcal{O}(1)$&$s\geq0$&$\mathcal{O}(1)$& $2\bar{r} + s +1$\\
      (C)& $\bar{r}\geq s$&$\mathcal{O}(1)$&$s \leq 0$ &$\mathcal{O}(1)$  & $2\bar{r}-s+1$\\
      (D)&$\bar{r}<0$&$\mathcal{O}(1)$&$s\geq \bar{r}$ &$\mathcal{O}(1)$& $1+s$ \\
      \hline
    \end{tabular}
    \label{tab:scaling_exponent}
\end{table}
To determine the time complexity, we parametrize
	\begin{equation}
		\gamma = \bar{\gamma}\,N^{-\bar r - 1}, \qquad 
		\kappa = \bar{\kappa}\,N^{-s},	
		\label{eq:scaling_rs}
	\end{equation}
with $\bar{\gamma},\bar{\kappa} = \mathcal{O}(1)$ and real exponents $\bar r,\,s$, which can vary from positive to negative values \footnote{Parametrization \eqref{eq:scaling_rs} is defined with respect to the energy $\epsilon_w$ of the target state, which is here set to unity. An equivalent parametrization would correspond to choosing $\gamma=\bar\gamma/N$, $\kappa=\bar\kappa N^{\bar r-s}$ and $\epsilon_w=N^{\bar r}$.}. Assuming $N\gg1$, parametrization \eqref{eq:scaling_rs} gives the time complexity $\tau(N)= \Theta(N^\alpha)$ with $\alpha$ real, over which the search asymptotically converges. 

\begin{figure}[t]
\includegraphics[width=\columnwidth]{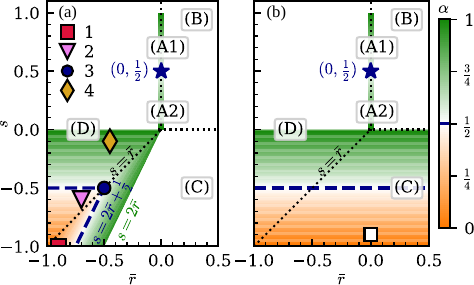}
\caption{Color plot of the exponent $\alpha$ as a function of  $\bar r$ and $s$ of the parametrization in Eq.\ \eqref{eq:scaling_rs} for (a) the monitored search and (b) for the full resetting protocol (with optimized resetting time $T$)  The value $\alpha=1/2$ of Grover's time complexity is indicated by the blue dashed line and by the blue star. White regions have exponent $\alpha>1$. Dotted lines separate different scaling regimes, labeled (B), (C), and (D), see Table~\ref{tab:scaling_exponent}. In the resetted search, in regime (C) the exponent of the time-complexity is now $\alpha=s+1$. It remains unchanged in regime (D). The symbols indicate the points studied numerically in Fig.\ \ref{fig:3}.} 
    \label{fig:2}
\end{figure}

Table~\ref{tab:scaling_exponent} summarizes the different regimes determining the scaling of $\alpha$, see also Fig.~\ref{fig:2}. Figure~\ref{fig:2}(a) shows the regimes in the $\bar r-s$ plane in which the sublinear scaling, $\alpha<1$, can be achieved, namely, when time complexity of the quantum algorithm could be more advantageous than that of the classical search algorithm. The green regions correspond to $\alpha\in (1/2,1)$, where the time complexity is better than classical but worse than Grover's time complexity. The shading from dark to light green indicates decreasing $\alpha$. These regions comprise the line at $\bar r=0$: The exceptional point (blue star) reaches $\alpha=1/2$ and is a local minimum along the axis. It separates the two intervals (A1) and (A2), where the exponent $\alpha$ increases moving away from $s=1/2$, till the algorithm reaches the time complexity of a classical search at $s=1$ and $s=0$, respectively. The behavior of the time complexity along this interval is reminiscent of noise-induced resonances in classical optimization, where convergence to an optimal solution is boosted at optimal amplitudes of the noise~\cite{Folz:2023}. Quantum advantage, namely, $\alpha<1$, is also found in the colored area at $\bar r,s<0$, where the tunneling amplitude and monitoring rate increase with $N$. Grover's time complexity is at the blue dashed line, while in the orange region $\alpha<1/2$. Here, the shading from light to dark orange indicates decreasing $\alpha$. At $\bar r=s=-1$ the time complexity changes from NP to P.

\emph{Quantum resetting.\textemdash}The resetting protocol repeats identical epochs of length $T$ (see Fig.~\ref{fig:1}) until a detector click occurs. We demand $\mathcal F_1 \ge 1-\varepsilon$ \eqref{eq:F1_main_text}, namely, $P(T)^m<\varepsilon$. This provides a lower bound to number of epochs: $m \ge \lceil{\tfrac{|\ln\varepsilon|}{|\ln P(T)|}}\rceil$.  We generally observe that there is an optimal reset time, $T>0$, which minimizes the search time. We verify whether resetting might also modify the time complexity through the scaling with $N$ of the search time $\tau_R=mT$, which is determined by the functional form of the no-click probability $P(T)$ at the end of an epoch. Note that at the exceptional point there is no advantage in resetting with respect to simply monitoring the dynamics. In fact, here $P(T)=e^{-\lambda T}$ with $\lambda$ a real and positive scalar, thus $P(T)^m=P(mT)$, see also~\cite{Chechkin:2018}. Away from the exceptional point, the reset time can be optimized to speed up the search \cite{roy2025causalitylocalisationuniversalitymonitored}. For instance, in regime (B), for $s>0$, choosing $T = N^{s}$ leads to a change of the time complexity from $\tau=\Theta( N^{2\bar r+s+1})$ to the more favorable scaling $\Theta(N^{s+1})$, see SM~\cite{SM2}. For $s<0$ choosing the scaling $T = N^{-|s|}$ has drastic effects: quantum advantage, $\alpha<1$, is now found in the whole plane at $s<0$, see Fig.\ \ref{fig:2}(b). 

\emph{Speed limits and query complexity.\textemdash}The analysis of the quantum search protocol shows that monitoring and resetting could lead to asymptotic convergence to the target with time complexity that scales sublinearly with $N$. Interestingly, at the optimal point $\gamma=1/N$ of the unitary search, adding a monitoring at rate $\kappa=\bar\kappa/\sqrt{N}$ warrants that the protocol converges asymptotically to the target with a characteristic timescale $\tau=\Theta(\sqrt{N})$. The time complexities found for monitoring rates $\kappa=\bar\kappa N^{|s|}$ can even change from NP to polynomial. This behavior is typically encountered for dynamics violating no-signaling principles~\cite{Bao:2016}. In our case, this apparent contradiction is solved by determining the scaling of the resources needed for implementing the measurement processes of Eq.\ \eqref{eq:Liouville}. 
In the SM we determine the query complexity for realizing the Lindbladian dynamics Eq.\ \eqref{eq:Liouville} on a quantum computer \cite{Cleve_2017, Ding_2024, Berry2007, berry2013gateefficientdiscretesimulationscontinuoustime}. 

\begin{figure}[t]
    \centering
    \includegraphics[width=\linewidth]{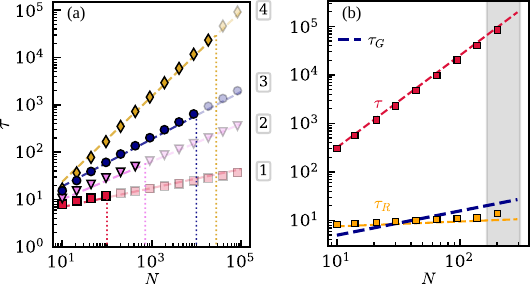}
    \caption{(a) Scaling of the time $\tau$ of the monitored search with the database size $N$. The symbols of the four curves correspond to the values of $\bar r, s$ of the symbols in Fig.~\ref{fig:2}(a). The dotted line marks $N_i^*$ (here for $\mathrm{d}t_0=0.01$) where the predictions of the model become invalid. (b) Comparison between the scaling of the search times $\tau$ (monitored search) and $\tau_R$ (resetting) as a function of $N$. The data is numerical and determined for the squared symbol of Fig.\,\ref{fig:2}(b). The shaded region demarcates the areas where $N>N^*$. The blue line gives $\tau_G=\Theta(\sqrt{N})$. In all plots, the underlying dashed lines show the analytic scaling, demonstrating quantitative agreement between numerical and analytical results. See~\cite{data_non_hermitian_search}.}
    \label{fig:3}
\end{figure}

We here discuss the scaling of the resource \cite{atia_fast-forwarding_2017} of the analog protocol  for $s\leq 0$, where monitoring is realized by alternating coherent evolution of duration $\mathrm{d}t$ with a projective measurement, for instance, by coupling the target site to a spectator qubit \cite{Singh:2023,Scholl2023,Wu2022,tao2024high} or to a resonator \cite{Anikeeva_2021,Morigi:2015}. The continuous description of Eq.\ \eqref{eq:Liouville} is valid when $\gamma N \mathrm{d}t \ll 1$, which bounds the monitoring rate: $\kappa\lesssim \gamma^2N^2 \mathrm{d}t$ \cite{Dhar2015QuantumModel}. The time step ${\rm d}t$, in turn, is bound from below by the physical time scale ${\rm d}t_0$, which depends on the energy cutoff and is independent of $N$. This leads to the relation $s\ge 2\bar r$. As a result, the search time of the monitored protocol is always bound by Grover scaling, $\alpha\ge 1/2$. Moreover, ${\rm d}t_0$ sets a lower bound to the resetting time, since $T\ge {\rm d}t$. Therefore, time complexities with $\alpha<1$ are only reached in the intervals (A1), (A2), and at the exceptional point.  Nevertheless, the relaxation in the regions at $s<0$ in Fig.~\ref{fig:2} can still be realized for databases with size $N\lesssim N^*$, where one could design ${\rm d}t={\rm d}t(N)$ such that the physical constraints are satisfied. The bound $N^\star$ is determined by the speed limit, ${\rm d}t_0={\rm d}t(N^*)$. Figure \ref{fig:3}(a) illustrates the scaling of the search time $\tau$ with $N$ for four representative points of diagram \ref{fig:2}(a), highlighting the bound $N^*$ above which the predictions of the model become invalid. Figure \ref{fig:3}(b) compares the scaling of the search time without and with resetting for a point at $s<0$, showing that the search can be significantly speed up for $N<N^*$.

\emph{Outlook.\textemdash}The time complexity of a quantum search can be tailored by means of non-unitary processes, such as the monitoring and a resetting rate, achieving asymptotic convergence and quantum advantage. The phase diagrams of Fig.\ \ref{fig:2} indicate the existence of measurement-induced dynamical phase transition \cite{Skinner:2019}, here in the time-complexity. The dynamics here discussed is amenable to generalization to many-body settings \cite{Puente_2024,solanki2025universalrelaxationspeedupopen} and
indicate a path to design rapid cooling protocols (see \cite{Zhan_etal2025_RapidQuantumGround,Allahverdyan_Petrosyan2022_DissipativeSearchUnstructured}): cooling is fastest for monitoring rates scaling with $N$, which can be realized by means of superradiant emission \cite{Periwal_etal2021_ProgrammableInteractionsEmergent,Jaeger:2022}. Here, the bound $N^*$ naturally emerges from the energy cutoff, which limits the validity of the underlying low-energy theoretical model \cite{Halati:2025,Halati:2025b}.

\emph{Acknowledgments.\textemdash}The authors acknowledge discussions with and helpful comments of David Gross, Shamik Gupta, Peter Orth, Adi Pick, Matteo Rizzi, and Tom Schmit. This work was funded by the German Ministry of Education and Research (BMBF, Project ``NiQ: Noise in Quantum Algorithms"), by the Deutsche Forschungsgemeinschaft (DFG, German Research Foundation) – Project-ID 429529648 – TRR 306 QuCoLiMa (``Quantum Cooperativity of Light and Matter''), and by the QuantERA II Programme (project ``QNet: Quantum transport, metastability, and neuromorphic applications in Quantum Networks"), which has received funding from the EU's Horizon 2020 research and innovation programme under Grant Agreement No.\ 101017733, as well as from the Deutsche Forschungsgemeinschaft DFG (Project ID 532771420).

\emph{Data availability.\textemdash}The data that supports the findings of this article are openly available at \cite{data_non_hermitian_search}.

\bibliography{reference}

\clearpage 
\onecolumngrid
\allowdisplaybreaks
\setcounter{equation}{0}
\setcounter{figure}{0}
\setcounter{table}{0}
\setcounter{section}{0}
\renewcommand{\theequation}{S\arabic{equation}}
\renewcommand{\thefigure}{S\arabic{figure}}
\renewcommand{\thetable}{S\arabic{table}}
\renewcommand{\thesection}{\Roman{section}}
\setcounter{secnumdepth}{3}               
\makeatletter
\renewcommand\@seccntformat[1]{\csname the#1\endcsname.\quad}
\makeatother

\setcounter{page}{1}
\renewcommand{\thepage}{S\arabic{page}}

\begin{center}
  \textbf{\Large Supplemental Material for ``Time complexity of a monitored quantum search with resetting''}\\[12pt]
  \normalsize Emma C. King \orcidlink{0000-0002-6696-3235},$^{1,*}$ Sayan Roy 
\orcidlink{0009-0005-5805-3791},$^{1,*}$ Francesco Mattiotti \orcidlink{0000-0002-2532-8876},$^{1}$ Maximilian Kiefer-Emmanouilidis \orcidlink{0009-0005-7065-3511},$^{2,3}$ \\Markus Bl\"aser \orcidlink{0000-0002-1750-9036},$^{4,5}$ and Giovanna Morigi \orcidlink{0000-0002-1946-3684}$^{1,5}$  \\[4pt]
    {\itshape \small
  $^{1}$Theoretical Physics,  Saarland University,  D-66123  Saarbr\"ucken,  Germany\\
  $^{2}$Embedded Intelligence, German Research Center for Artificial Intelligence (DFKI), D-67663 Kaiserslautern, Germany\\
  $^{3}$Department of Computer Science and Research Initiative QC-AI, RPTU Kaiserslautern-Landau, D-67663 Kaiserslautern, Germany\\
  $^{4}$Computer Science, Saarland University, D-66123 Saarbr\"ucken, Germany\\
  $^{5}$Center for Quantum Technologies (QuTe), Saarland University, Campus, 66123 Saarbr\"ucken, Germany}
  
  {\small (Dated: January 28, 2026)}
\end{center}

\twocolumngrid
\tableofcontents

\begin{figure}[!b]
\centering
\begin{minipage}{0.95\columnwidth}
\raggedright \noindent\rule{0.2\linewidth}{0.4pt}\\ \vspace{0.5cm}
\footnotesize
\raggedright $^*$ These authors contributed equally to this work.
\end{minipage}
\end{figure}

\section{Derivation of $\hat H_{\rm{eff}}^{(s)}$}
\label{app:1}
The Lindblad master equation for the system's density matrix reads as (in units where $\hbar=1$)
\begin{equation}
    \dot{\hat{\varrho}} = - i [\hat{H}, \hat{\varrho}] + \sum_i k_i (\hat{J}_i \hat\varrho \hat{J}_i^\dagger - \frac{1}{2} \{\hat{J}_i^\dagger \hat{J}_i, \hat{\varrho} \})\,,
    \label{eq:master_equation1}
\end{equation}
where $[\hat{a},\hat{b}] = \hat{a}\hat{b} -\hat{b}\hat{a}$ denotes the commutator and  $\{\hat{a},\hat{b}\} = \hat{a}\hat{b} + \hat{b}\hat{a}$ denotes the anti-commutator. $\hat{H}$ is the system Hamiltonian that governs the unitary dynamics. $\{ \hat{J}_i \}$ is the set of jump operators that governs the dissipative part of the dynamics. $k_i \geq 0$ denotes the damping rates. The entire equation can be written in a superoperator form like $\dot{\hat{\varrho}} = \mathcal{L} \hat{\varrho}$, where $\mathcal{L}$ is the Liouvillian superoperator. One may also view this as a completely positive trace preserving (CPTP) map (quantum channel) $\Lambda_t$ which maps the initial density matrix $\hat{\varrho}_0$ to the final density matrix $\hat{\varrho}_t$. Mathematically, $\hat{\varrho}_t =  \Lambda_t [\hat{\varrho}_0] = e^{t\mathcal{L}}\hat{\varrho}_0$. 

In our setup, we have an ancillary ``detector'' which monitors the target vertex $\ket{w}$.  We model the detector by a two-level system, such that states $|0_\mathrm{D}\rangle$ and $|1_\mathrm{D}\rangle$ correspond to no-click and one click, respectively. The detection process is described by an irreversible decay from the state $|w,0_\mathrm{D}\rangle$ to the state $|w,1_\mathrm{D}\rangle$. We model this process via a single jump operator $J_i = \ket{w, 1_\mathrm{D}}\bra{w, 0_\mathrm{D}}$ with a rate governed by $\kappa$. Eq.~\eqref{eq:master_equation1} can now be written as 
\begin{eqnarray}  
        \dot{\hat{\varrho}} = \mathcal{L}\hat{\varrho} &=& - i [\hat{H}, \hat{\varrho}] + 2 \kappa (\ket{w, 1_\mathrm{D}}\bra{w, 0_\mathrm{D}} \hat{\varrho} \ket{w, 0_\mathrm{D}}\bra{w, 1_\mathrm{D}} \nonumber \\  &&- \kappa \{\ket{w, 0_\mathrm{D}}\bra{w,0_\mathrm{D}} , \hat{\varrho} \})\,, \nonumber \\
        &=& - i \left(\hat{H}_{\rm eff} \hat{\varrho} -  \hat{\varrho} \hat{H}_{\rm eff}^\dagger \right) \nonumber \\ &&+ 2 \kappa \ket{w, 1_\mathrm{D}}\bra{w, 0_\mathrm{D}} \hat{\varrho} \ket{w, 0_\mathrm{D}}\bra{w, 1_\mathrm{D}}\,,
     \label{eq:master_equation2}
\end{eqnarray}
where 
\begin{equation}
   \hat{H} = - \gamma N \ket{s}\bra{s} - \epsilon_w\ket{w}\bra{w}
   \label{eq:bare_hamiltonian}
\end{equation}
 is the bare system  Hamiltonian for $\kappa = 0$ and $\ket{s}$ is the uniform superposition state. The operator $\hat{H}_{\rm eff} = \hat{H}^{(s)}_{\rm eff} \otimes \ket{0_\mathrm{D}}\bra{0_\mathrm{D}}$ where $\hat{H}_{\rm eff}^{(s)} = \hat{H} -i \kappa \ket{w}\bra{w}$ is the effective non-Hermitian Hamiltonian that governs the dynamics of the system. 

\section{Derivation of the no-click probability}
\label{app:2}
For a generic non-Hermitian operator $\hat{A} \neq \hat{A}^\dagger$, the right and left eigenvectors need not coincide. Assuming a non-degenerate spectrum, the operator is diagonalizable and admits a complete biorthogonal set of eigenvectors (otherwise one must work with Jordan blocks of generalized eigenvectors)~\cite{morse_methods_1953,brody_biorthogonal_2014}. In this work, we consider a non-Hermitian search Hamiltonian $\hat{H}_{\rm eff}^{(s)}$ that is diagonalizable with a complete biorthogonal set of right and left eigenvectors $\left\{\,\ket{ \lambda_j^{\prime}},\bra{\bar{\lambda}_j^{\prime}}\,\right\}$ (both associated with the same set of eigenvalues $\lambda_j$~\cite{morse_methods_1953}) satisfying
\begin{eqnarray}
	\hat{H}_{\rm eff}^{(s)} \ket{\lambda_j^{\prime}} &=& \lambda_j\ket{\lambda_j^{\prime}}\,, \label{eq:eig_eq_right}\\
	\bra{\bar{\lambda}_j^{\prime}} \hat{H}_{\rm eff}^{(s)} &=& \bra{\bar{\lambda}_j^{\prime}} \lambda_j\,.
\label{eq:eig_eq_left}
\end{eqnarray}
Here, the eigenvectors are not yet normalized; we fix the normalization below. Taking adjoints of Eqs.~\eqref{eq:eig_eq_right}–\eqref{eq:eig_eq_left} gives the corresponding relations for $\hat{H}_{\rm eff}^{(s) \,\dagger}$,
\begin{eqnarray}
    \bra{\lambda_j^\prime} \hat{H}_{\rm eff}^{(s) \,\dagger} &=& \bra{\lambda_j^\prime} \lambda_j^*\,, \label{eq:eig_eq_right_adj}\\
  \hat{H}_{\rm eff}^{(s) \,\dagger} \ket{\bar{\lambda}_j^\prime} &=& \lambda_j^* \ket{\bar{\lambda}_j^\prime}\,.
     \label{eq:eig_eq_left_adj}
\end{eqnarray}
To fix the normalization, one needs to impose the biorthonormality condition $\langle \bar{\lambda}_j \vert \lambda_k\rangle = \delta_{j,k}$, where $\ket{\lambda_k}$, $\bra{\bar{\lambda}_j}$ are normalized vectors. One convenient choice of normalization is
\begin{equation}\label{eq:eigenvector_normalization}
  \ket{\lambda_j} =\frac{1}{\sqrt{\langle \bar{\lambda}_j^\prime \vert \lambda_j^\prime \rangle}}  \ket{\lambda_j^\prime}\,, \quad
  \bra{\bar{\lambda}_j} = \frac{1}{\sqrt{\langle \bar{\lambda}_j^\prime \vert \lambda_j^\prime \rangle}}  \bra{\bar{\lambda}_j^\prime}\,.
\end{equation}
With this normalization, the completeness relation reads
\begin{equation}\label{eq:completeness}
    \sum_j \ket{\lambda_j}\bra{\bar{\lambda}_j} =  \mathds{1} \,.
\end{equation}
The no-click probability is given by the norm of the time-evolved state $\ket{\psi(t)}$,
\begin{equation}
P(t)=\langle \psi(t)\vert \psi(t) \rangle =\bra{s}\,e^{i\, \hat{H}_{\rm eff}^{(s) \,\dagger}\, t}e^{-i\, \hat{H}_{\rm eff}^{(s)}\, t}\,\ket{s}\,,
\end{equation}
where $\ket{s}$ is the initial uniform superposition state. Inserting the completeness relation (Eq.~\eqref{eq:completeness}) gives 
\begin{align}
P(t) &=\bra{s}\,\mathds 1\,e^{i\, \hat{H}_{\rm eff}^{(s) \,\dagger}\, t}e^{-i\, \hat{H}_{\rm eff}^{(s)}\, t}\,\mathds 1\,\ket{s}\,, \nonumber\\
&=\sum_{j,k=+,-} \langle s \vert \bar{\lambda}_j \rangle \langle \lambda_j \vert e^{i\, \hat{H}_{\rm eff}^{(s) \,\dagger}\, t}e^{-i\, \hat{H}_{\rm eff}^{(s)}\, t} \vert \lambda_k \rangle \langle \bar{\lambda}_k\vert s \rangle\,, \\
&=\sum_{j,k =+,-} e^{-i(\lambda_k-\lambda_j^*)t} b_{jk} \langle s \vert \bar{\lambda}_j \rangle \langle \bar{\lambda}_k \vert s \rangle\,,
\label{eq:surv_prob_appendix}
\end{align}
where $b_{jk} = \langle \lambda_j \vert \lambda_k \rangle $  and in the last line we have used Eq.~\eqref{eq:eig_eq_right} and Eq.~\eqref{eq:eig_eq_right_adj}. Eq.~\eqref{eq:surv_prob_appendix} is the same as the Eq.~(8) of the main text and $+,-$ represents the eigenvalue $\lambda_+, \lambda_-$ respectively (See Eq.~(7) of the main text).

For $j=k$,
\begin{equation}
P_{(j=k)}(t)=\sum_{j = +,-} e^{2 \text{Im}(\lambda_j) t} \,  b_{jj} \,  \vert \langle s \vert \bar{\lambda_j} \rangle \vert^2 \,,
\end{equation}
and for $j \neq k$, pairing $(+,-)$ with $(-,+)$  terms yields a real part,
\begin{equation}
P_{(j\neq k)}(t) = 2 \text{Re} \left[e^{-i(\lambda_+ - \lambda_-^*)t} \langle s \vert \bar{\lambda}_{-} \rangle \langle \bar{\lambda}_+ \vert s \rangle\,\, b_{- +}\right]\,.
\end{equation}
Combining $P_{(j=k)}(t)$ and $P_{(j\neq k)}(t)$, one obtains the expression of no-click probability as 
\begin{align}
P(t) =&\; e^{2 \, \text{Im}(\lambda_+)\, t} \vert \langle s \vert \bar{\lambda}_+ \rangle \vert^2 \,\, b_{++} + e^{2 \, \text{Im}(\lambda_-)\, t} \vert \langle s \vert \bar{\lambda}_- \rangle \vert^2 \,\, b_{--}  \nonumber \\ 
    &+ 2 \text{Re} \left[e^{-i(\lambda_+ - \lambda_-^*)t} \langle s \vert \bar{\lambda}_- \rangle \langle \bar{\lambda}_+ \vert s \rangle\,\, b_{- +} \right].
\label{eq:surv_prob_expand}
\end{align}
Next we define the overlaps which will be used in later sections: $O_+ \equiv  \vert \langle s \vert \bar{\lambda}_+ \rangle \vert^2 \, b_{++}$, $O_- \equiv \vert \langle s \vert \bar{\lambda}_- \rangle \vert^2\, b_{--} $ and $O_\pm \equiv \langle s \vert \bar{\lambda}_- \rangle \langle \bar{\lambda}_+ \vert s \rangle \,\, b_{-+} $. With this notation, the no-click probability takes the simplified form 
\begin{align}
P(t) =& e^{2 \, \text{Im}(\lambda_+)\, t} O_+ + e^{2 \, \text{Im}(\lambda_-)\, t} O_-
\nonumber\\ &+ 2 \text{Re} \left[e^{-i(\lambda_+ - \lambda_-^*)t} O_\pm \right].
\label{eq:surv_prob_simplified_appendix}
\end{align}

\section{Search fidelities in presence of periodic resetting}
\label{app:3}
We consider deterministic periodic resets at times $t = T,2T,\ldots mT,\ldots$. Throughout, $P(t)$  denotes the no-click (survival) probability of the detector up to time $t$ starting from the initial state $\hat\varrho_0 = \ket{s,0_\mathrm{D}} \bra{s, 0_\mathrm{D}} = \ket{s}\bra{s} \otimes \vert  0_\mathrm{D}\rangle\langle{0_\mathrm{D}\vert} $, where $\ket{s}$ is the initial uniform superposition state and $\ket{0_\mathrm{D}}$ is the no-click state of the detector. Between resets the system evolves under the conditional (no-click) non-Hermitian dynamics generated by $\hat{H}_{\rm eff}^{(s)}$ as shown in Sec.~\ref{app:1}.

As discussed in the main text, for $\kappa>0$, we quantify the dynamics by search fidelity which is the total probability that the walker occupies the marked site $\ket{w}$ together with detector state $j_\mathrm{D} \in \{0,1\}$: $\mathcal F_{j=0,1}(t)={\rm Tr}\{|w,j_\mathrm{D}\rangle\langle w, j_\mathrm{D}|\hat{\varrho}_t\}$. These quantities measure the probability that the walker has reached the target conditioned on no detection ($\mathcal F_0$) or a detector click ($\mathcal F_1$). 

Without loss of generality, let us choose  $t\in(t_m,t_{m+1})$ with $t_m=mT$. For convenience, define $P_m\coloneqq P(T)^m$ and $\Delta\coloneqq t-t_m\in[0,T)$. Immediately after one reset to the initial state $\hat \varrho_0$ at time $t_1 = T^+$, the density matrix reads
\begin{eqnarray}
    \hat{\varrho}_{t_1} = (1 - P_1) \vert w, 1_\mathrm{D} \rangle\langle w, 1_\mathrm{D} \vert + P_1\, \hat\varrho_0\,.
\end{eqnarray}
Iterating once more, and using that the ``surviving part'' is re-prepared to $\hat\varrho_0$ at each reset, one finds at $t_2=2T^+$
\begin{eqnarray}
    \hat\varrho_{t_2} &=& (1 - P_1) \vert w, 1_\mathrm{D} \rangle\langle w, 1_\mathrm{D} \vert \nonumber \\
    &+& P_1(1 - P_1) \vert w, 1_\mathrm{D} \rangle\langle w, 1_\mathrm{D} \vert  + P_2\, \hat\varrho_0\,.
\end{eqnarray}
By induction, just after the $m^{\rm th}$ reset, the density matrix reads
\begin{eqnarray}
    \hat \varrho_{t_m} &=& (1 - P_1)\left[\sum_{n = 0}^{m-1} P_n\right] \vert w,1_\mathrm{D} \rangle\langle w, 1_\mathrm{D} \vert + P_m \hat\varrho_0\,, \nonumber \\
    &=& (1 - P_m)\vert w, 1_\mathrm{D} \rangle\langle w, 1_\mathrm{D} \vert + P_m\, \hat \varrho_0\,,
\end{eqnarray}
which makes explicit the renewal structure induced by periodic resetting. Now at a generic time $t$ between the $m^{\rm th}$ and $(m+1)^{\rm th}$ reset, the density matrix reads
\begin{align}
    \hat\varrho_{t} =&\, (1 - P_m)\vert w, 1_\mathrm{D} \rangle\langle w, 1_\mathrm{D} \vert + P_m(1 - P(\Delta) )\vert w, 1_\mathrm{D} \rangle w, \langle 1_\mathrm{D} \vert \nonumber\\&+ P_m e^{-i \hat{H}_{\rm eff}^{(s)} \Delta}\,\hat\rho_0 \,e^{i \hat{H}^{(s),\,\dagger}_{\rm eff} \Delta}\,, \nonumber \\
    =&\,(1 - P_m P(\Delta))\vert w, 1_\mathrm{D} \rangle\langle w,1_\mathrm{D} \vert \nonumber\\&+ P_m\, e^{-i \hat{H}_{\rm eff}^{(s)} \Delta}\,\hat\rho_0 \,e^{i \hat{H}^{(s),\,\dagger}_{\rm eff} \Delta}\,.
\end{align}
In the first line of the equation, the first term gives the probability of detection up to time $t = t_m$, the second term denotes no-click until time $t = t_m$ and detection in the time interval $t_m + \Delta$, whereas the third term indicates the probability of no-click till time $t$. The fidelities are thus obtained as follows
\begin{eqnarray}
\label{eq:app:fidelities}
 \mathcal{F}_0 &=& {\rm Tr}\{|w,0_\mathrm{D}\rangle\langle w, 0_\mathrm{D}|\hat\varrho_t\} = P_m \vert \langle w \vert e^{-i \hat{H}_{\rm eff}^{(s)} \Delta} \vert s \rangle \vert^2\,,  \nonumber  \\
  \mathcal{F}_1 &=& {\rm Tr}\{|w,1_\mathrm{D}\rangle\langle w, 1_\mathrm{D}|\hat\varrho_t\} = 1 - P_m P(\Delta)\,.
\end{eqnarray}

\section{Competing fidelity contributions}\label{app:4}
The search fidelity in the presence of periodic resetting comprises two \textit{competing} contributions $\mathcal F_0$, $\mathcal F_1$; see Sec.~\ref{app:3} and Eq.~\eqref{eq:app:fidelities}. In this section we demonstrate how maximizing the one fidelity contribution necessarily and unavoidably minimizes the other. 

Consistent with the notation of Sec.~\ref{app:3}, let $t\in(t_m,t_{m+1})$ with $t_m=mT$, $\Delta  = t - t_m$ and consider the no-click probability $P(t)$ as defined in Eq.~\eqref{eq:surv_prob_simplified_appendix}. With the walker initialized in $\ket{s}$, the two fidelity contributions are given by Eq.~\eqref{eq:app:fidelities}.
The objective is to maximize the total fidelity $\mathcal F(t)= \mathcal F_0(t) + \mathcal F_1(t)$. As mentioned in the main text, this requires two somewhat orthogonal optimization strategies. We now demonstrate that $\mathcal F_0(t)$ and $\mathcal F_1(t)$ cannot be simultaneously optimized, since the two fidelity contributions respond \emph{oppositely} to parameter changes that increase/decrease the no-click probability $P(T)$. 

To proceed, we introduce constants $c_0,\,c_1\in[0,1]$ with $c_0+c_1=1$ and nonnegative errors $\varepsilon_0(t),\,\varepsilon_1(t)\geq0$, such that 
\begin{equation}\label{eq:F0_F1_with_errors}
    \mathcal F_0(t)=c_0-\varepsilon_0(t),\qquad
	\mathcal F_1(t)=c_1-\varepsilon_1(t)\,.
\end{equation}
The total error is
\begin{equation}
	\varepsilon_{\rm tot}(t)=\varepsilon_0(t)+\varepsilon_1(t)=1-\mathcal F(t).
\end{equation}
Minimizing $\varepsilon_{\rm tot}$ is thus equivalent to maximizing $\mathcal F(t)=\mathcal F_0(t)+\mathcal F_1(t)$. The maximization of total fidelity leads to a clear trade-off, since the two terms $P_m\,|\langle w|e^{-i\hat{H}_{\mathrm{eff}}^{(s)}\Delta}|s\rangle|^2$ and $1-P_m\, P(\Delta)$ react monotonically and oppositely to the no-click probability $P(t)$. Fixing $m$, we have 
\begin{equation}
  \frac{\partial \mathcal F_1}{\partial P_m} =-P(\Delta) \le 0\,,\qquad 
  \frac{\partial \mathcal F_1}{\partial P(\Delta)} =-P_m \le 0\,,
\end{equation}
while for $\mathcal F_0$,
\begin{align}
	\frac{\partial\mathcal F_0}{\partial P_m} = |\langle w|e^{-i\hat{H}_{\mathrm{eff}}^{(s)}\Delta}|s\rangle|^2 \ge 0\,, \\
	\frac{\partial\mathcal F_0}{\partial\,|\langle w|e^{-i\hat{H}_{\mathrm{eff}}^{(s)}\Delta}|s\rangle|^2 }  = P_m \ge 0\,,
\end{align}
Consequently, independent of the parameter choice of $(\gamma,\,\kappa)$, increasing $P_m$ increases $\mathcal F_0$ (thus decreasing the error $\varepsilon_0$) while simultaneously decreasing $\mathcal F_1$ increases the error $\varepsilon_1$. The converse also holds.

Now note that
\begin{equation}\label{eq:p_le_S}
	|\langle w|e^{-i\hat{H}_{\mathrm{eff}}^{(s)}\Delta}|s\rangle|^2\leq  \langle s| e^{i\hat{H}^{(s)\,\dagger}_{\rm eff}\Delta} e^{-i\hat{H}_{\mathrm{eff}}^{(s)}\Delta}|s\rangle = P(\Delta)\,.
\end{equation} 
The behavior of the total fidelity is thus given by
\begin{eqnarray}
	\frac{\partial (\mathcal{F}_0 + \mathcal{F}_1) }{\partial P_m} = |\langle w|e^{-i\hat{H}_{\mathrm{eff}}^{(s)}\Delta}|s\rangle|^2 - P(\Delta) \leq 0\,,
\end{eqnarray}
where the inequality follows from the Cauchy-Schwarz inequality \eqref{eq:p_le_S}. Since the total fidelity is decreasing in $P_m$, the optimal strategy for $\kappa > 0$ is to choose parameters such that the no-click probability is minimized. In the ideal limiting case $P_m \to 0$, one has $\mathcal F_0(t)\to 0$ and $\mathcal F(t) = \mathcal F_1(t) \to 1$. By contrast, for $\kappa = 0$, $\mathcal F_1$ is trivially $0$ and $P(T)$ is maximized to $1$, so the optimal strategy is to maximize $\mathcal F_0$, thereby recovering the usual unitary spatial search setting~\cite{Childs_2004}.

\section{Asymptotic scaling of the search time $\tau$} \label{app:5}
This section provides detailed calculations on the asymptotic scaling of the search time $\tau$, defined as the first time at which the fidelity $\mathcal F(t)$ reaches a threshold value $\mathcal F_{\rm max}=1-\varepsilon$. From the results we identify three scaling regimes, as well as the behavior at the exceptional point, which are reported in the main text. See also Fig.~2.

We restrict our analysis to the case where $\kappa>0$. In accordance with Sec.~\ref{app:4}, the fidelity is then maximized by minimizing the no-click probability, i.e.\ we focus on the fidelity contribution $\mathcal F_1$ \eqref{eq:app:fidelities}. It follows that the scaling of the search time $\tau$ is set by the timescale governing the decay of the no-click probability. Referring back to the probability expression in Eq.~\eqref{eq:surv_prob_simplified_appendix}, the first two exponential terms represent incoherent decays with rates $2\,|\mathrm{Im}\,\lambda_\pm|$, while the third term exhibits oscillatory behavior and describes the interference between the two eigenmodes. The oscillations have frequency $\Omega=\mathrm{Re}\,(\lambda_+-\lambda_-)$ and decay at rate $\Gamma_{\rm int}=\mathrm{Im}\,(\lambda_++\lambda_-)$. The \textit{decay envelope} of $P(t)$ is set by the slower of the two exponentials, $e^{2\,\mathrm{Im}(\lambda_+)t}$ and $e^{2\,\mathrm{Im}(\lambda_-)t}$. We therefore relate the timescale of the search to the inverse of the imaginary part of the effective Hamiltonian eigenvalues as
\begin{equation}
	\tau=\Theta\left( \frac{1}{\min\{|\mathrm{Im}\,\lambda_+|,|\mathrm{Im}\,\lambda_-|\}} \right).
	\label{eq:Tdecaydef0}
\end{equation}
The scaling behavior of the eigenvalues $\lambda_\pm$ with $N$ directly imprints on the search time $\tau$. We proceed in Sec.~\ref{app:ss:asymptotics_evs} by deriving the asymptotics of the eigenvalues.

\subsection{Asymptotic scaling of eigenvalues}\label{app:ss:asymptotics_evs}
\begin{figure}[t]
    \centering
    \includegraphics[width=\linewidth]{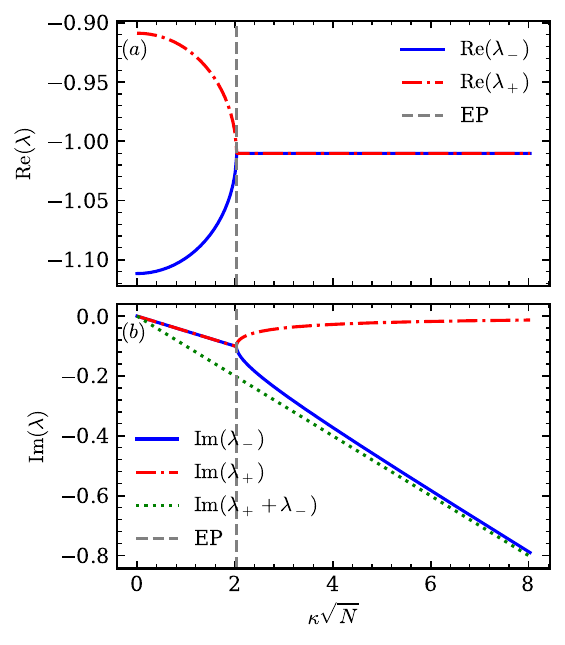}
    \caption{Dependence of the two eigenvalues $\lambda_{\pm}$ on the scaled monitoring rate $\kappa\sqrt{N}$ with $\gamma=\gamma^{\mathrm{EP}}$. \textbf{(a)} Real parts $\mathrm{Re}(\lambda_{\pm})$. \textbf{(b)} Imaginary parts $\mathrm{Im}(\lambda_{\pm})$. The vertical dashed line at $\kappa=\kappa^{\mathrm{EP}}\approx2/\sqrt{N}$ marks the exceptional point where the eigenvalues coalesce; for larger $\kappa$ they split along the imaginary axis. We take $N=100$.}
    \label{fig:EPs_fixed_gamma}
\end{figure}

\begin{figure}[t]
    \centering
    \includegraphics[width=\linewidth]{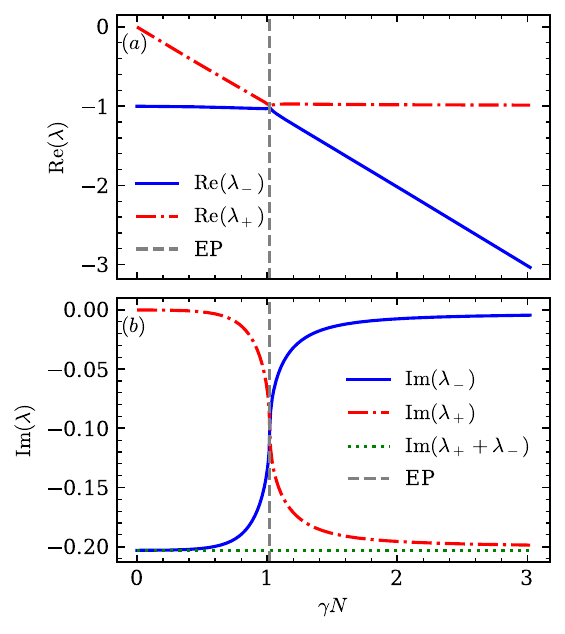}
    \caption{Dependence of the two eigenvalues $\lambda_{\pm}$ on the scaled hopping rate $\gamma N$ with $\kappa=\kappa^{\mathrm{EP}}$. \textbf{(a)} Real parts $\mathrm{Re}(\lambda_{\pm})$. \textbf{(b)} Imaginary parts $\mathrm{Im}(\lambda_{\pm})$. The vertical dashed line at $\gamma=\gamma^{\mathrm{EP}}\approx1/N$ marks the exceptional point where the eigenvalues coalesce. We take $N=100$.}
    \label{fig:EPs_fixed_kappa}
\end{figure}

Recall that $\hat{H}_{\rm eff}^{(s)} = \hat{H} -i \kappa \ket{w}\bra{w}$ with $\hat{H} = - \gamma N\hat{L} - \epsilon_w\ket{w}\bra{w}$. Consistent with the main text, we choose $\epsilon_w=1$. Writing the Hamiltonian $\hat{H}_{\rm eff}^{(s)}$ in the two-dimensional subspace of $\{\ket{w}, \ket{r_\perp} \equiv  1/\sqrt{N-1}\,\,\sum_{j=1,\,j\neq w}^{N} \lvert j \rangle\, \}$, we define the eigenvalues of the effective Hamiltonian, parametrized as $\hat H_{\rm eff}^{(s)} = \begin{pmatrix} a & b\\ b & d \end{pmatrix}$, as
\begin{equation}
	\lambda_\pm = \frac{a+d}{2}\pm
	\sqrt{\frac{(a-d)^2}{4}+b^2}\,,
	\label{eq:lpm-exact}
\end{equation}
where
\begin{align}\label{eq:a_b_d_definition}
	a&=-(\gamma+1+i\kappa)\,,\nonumber\\
	b&=-\gamma\sqrt{N-1}\,,\\
	d&=-\gamma(N-1)\,,\nonumber
\end{align}
with real, positive parameters $\gamma,\kappa$ and integer $N\ge1$. Define the discriminant $D\equiv \sqrt{\frac{(a-d)^2}{4}+b^2}$. An exceptional point (EP) of order two---a singularity in the parameter space of the non-Hermitian system described by $\hat{H}_{\rm eff}^{(s)}$ where the two eigenvalues \eqref{eq:lpm-exact} and their corresponding eigenvectors coalesce---occurs when $D=0$. Directly from the discriminant $D$ we find the parameter values $\gamma^{\rm EP}$ and $\kappa^{\rm EP}$ corresponding to the EP as
\begin{align}
	\gamma(N-2)-1 =0\quad&\Rightarrow\quad \gamma^{\rm EP}=\frac{1}{N-2},\nonumber\\
	-\kappa = \pm 2\gamma\sqrt{N-1}\quad&\Rightarrow\quad \kappa^{\rm EP}=\mp\frac{2\sqrt{N-1}}{N-2}.
\end{align}
Since we fix $\kappa>0$, we have a single exceptional point in parameter space at $(\gamma,\kappa) = (\gamma^{\rm EP},2\sqrt{N-1}/(N-2))$. This point separates regimes in which the eigenvalues behave very differently; see Figs.~\ref{fig:EPs_fixed_gamma} and \ref{fig:EPs_fixed_kappa}. 

In particular, fixing $\gamma=\gamma^{\rm EP}$ and for $\kappa<\kappa^{\rm EP}$, the two exponentials in the no-click probability decay at the same rate since $\mathrm{Im}(\lambda_+)=\mathrm{Im}(\lambda_-)$, while the third term decays more rapidly and oscillates at a frequency $\Omega=\mathrm{Re}(\lambda_+-\lambda_-)\neq 0$. In contrast, if $\kappa>\kappa^{\rm EP}$, the $\mathrm{Im}(\lambda_+)$ will dictate the timescale of decay of $P(t)$, see Eq.~\eqref{eq:Tdecaydef} and Fig.~\ref{fig:EPs_fixed_gamma}, and there will be no oscillations in the no-click probability since $\Omega=0$.

Similarly, we can identify different behaviors of the eigenvalues for fixed $\kappa=\kappa^{\rm EP}$. Independent of the value of $\gamma$, the interference term exhibits oscillations; see Fig.~\ref{fig:EPs_fixed_kappa}(a). If $\gamma<\gamma^{\rm EP}$, the search time is set by $\mathrm{Im}(\lambda_+)$ according to Eq.~\eqref{eq:Tdecaydef}. When $\gamma>\gamma^{\rm EP}$, we observe that the timescale is now determined by $\mathrm{Im}(\lambda_-)$. The transition in the imaginary parts of the eigenvalues of Fig.~\ref{fig:EPs_fixed_kappa}(b) becomes sharper with increasing $N$, and demonstrates that the search performance, determined by the search time scaling to acquire a fixed fidelity value, will deteriorate rapidly when moving away from the point $\gamma\approx\gamma^{\rm EP}$. More precisely, the dominant term of the no-click probability \eqref{eq:surv_prob_simplified_appendix} decays slowly at rate $\Theta( \min\{|\mathrm{Im}\,\lambda_+|,|\mathrm{Im}\,\lambda_-|\})$ and is large, since the corresponding overlap $O_i\approx1$, $i\in\{+,-\}$ (see Sec.~\ref{app:ss:overlaps} for a detailed discussion on the order of magnitude and scaling of the overlaps).

To further develop this understanding, we now describe several scaling regimes. All results, summarized in Table I of the main text, follow from controlled expansions of the square root \eqref{eq:lpm-exact} in the two asymptotic regimes $|b|\ll |a-d|$ and $|b|\gg |a-d|$, see Eq.~\eqref{eq:a_b_d_definition}. The crossover, where $|b|\sim |a-d|$, needs to be treated separately. For compactness, we write $\delta=(a-d)/2$ and $\bar a =(a+d)/2$. Moreover, we introduce the notation 
\begin{align}\label{eq:app:lambda_s_lambda_f}
    |\Im\lambda_s|&\equiv \min\{|\mathrm{Im}\,\lambda_+|,|\mathrm{Im}\,\lambda_-|\}\,,\\
    |\Im\lambda_f|&\equiv \max\{|\mathrm{Im}\,\lambda_+|,|\mathrm{Im}\,\lambda_-|\}\,,\label{eq:app:lambda_f}
\end{align}
where subscripts $s,\,f$ are used to denote the \textit{slow} and \textit{fast} decaying eigenmodes of the Hamiltonian. With this definition, the timescale of the search (Eq.~\ref{eq:Tdecaydef0}) is now given by 
\begin{equation}
	\tau=\Theta\left( \frac{1}{\min\{|\mathrm{Im}\,\lambda_+|,|\mathrm{Im}\,\lambda_-|\}} \right) = \Theta\left(\frac{1}{\vert\mathrm{Im}\,\lambda_s\vert}\right).
	\label{eq:Tdecaydef}
\end{equation}

Focusing first on the ``weak-mixing'' regime, $|b|\ll |a-d|$ (i.e.\ $|b/\delta|\ll1$), we view $b$ as a perturbation and a Taylor expansion of the root yields
\begin{equation}
	\sqrt{\delta^2+b^2} = \delta\Bigl(1 + \tfrac{b^2}{2\delta^2} +\mathcal{O} \left (\tfrac{b^4}{\delta^4}\right) \Bigr).
\end{equation}
Consequently, the nondegenerate perturbation expansion gives
\begin{equation}\label{eq:weak-mixing}
	\begin{aligned}
		\lambda_{f} &= a + \dfrac{b^2}{a-d} + \mathcal{O}\left(\dfrac{b^4}{(a-d)^3}\right),\\
		\lambda_{s} &= d - \dfrac{b^2}{a-d} + \mathcal{O}\left(\dfrac{b^4}{(a-d)^3}\right).
	\end{aligned}
\end{equation}
Note that the fast and slow eigenmode are defined according to larger imaginary part, see Eq.~\ref{eq:app:lambda_s_lambda_f} and Eq.~\ref{eq:app:lambda_f}. The fast and slow eigenvalue branches $\lambda_f$ and $\lambda_s$ are continuously connected to the unperturbed diagonals $a$ and $d$ of the effective Hamiltonian when $b\to0$. Crucially, $d\in \mathbb{R}$ while $a$ carries an imaginary part $\Im a \neq 0$ for $\kappa\neq0$, inferring that the \textit{slow} mode $\lambda_s$ associated with $d$ acquires a \textit{small} decay rate only through its coupling to the damped (fast) mode $\lambda_f$. This induced damping appears at second order in the mixing amplitude as $\Im\lambda_s \sim -\Im[b^2/(a-d)]$, and will set the timescale for the no-click probability decay; recall Eqs.~\eqref{eq:Tdecaydef} and \eqref{eq:app:lambda_s_lambda_f}. 
	
In the ``strong-mixing" regime where $|b|\gg |a-d|$, we expand in $\delta/b$ with $|\delta/b|\ll 1$:
\begin{equation}
	\sqrt{b^2+\delta^2} = b\Bigl(1 + \tfrac{\delta^2}{2b^2} + \mathcal{O} \left (\tfrac{\delta^4}{b^4}\right)\Bigr).
\end{equation}
It follows directly that the eigenvalues take the form
\begin{equation}\label{eq:strong-mixing}
	\lambda_{f,s} = \bar a \pm b \pm \frac{\delta^2}{2b} + \mathcal{O}\!\Bigl(\frac{\delta^4}{b^3}\Bigr).
\end{equation}
In this case, the two modes are strongly hybridized and split by $\sim 2|b|$. Since $\Im a \neq 0$ for $\kappa\neq0$, both modes share the leading-order imaginary part, and $\delta$ produces subleading corrections. The crossover $|b|\sim |a-d|$ cannot be handled perturbatively and is treated on a case-by-case basis. In fact, this regime is special in that it includes exceptional-point behavior when the discriminant $D$ vanishes and is characterized by non-analytic crossovers in search-time scaling.
	
Using the above expansions, we now derive the explicit asymptotic forms of the search-time scaling $\tau=\Theta(N^{\alpha})$ appearing in Table I (see main text). The hopping and monitoring rates are parametrized as in the main text:
\begin{equation}
	\gamma = \bar{\gamma}\,N^{-\bar r-1},\qquad
	\kappa = \bar\kappa\,N^{-s},
	\qquad \bar\gamma,\bar\kappa\in \mathbb{R}^+,
	\label{eq:scalings_rs}
\end{equation}
with $\bar\gamma,\,\bar\kappa = \mathcal{O}(1)$ and exponents $\bar r,s\in\mathbb{R}$ controlling the strength of coherent hopping and monitoring, respectively. Henceforth, for ease of notation, we perform derivations with a modified hopping exponent $r\equiv \bar r +1$. Using definition \eqref{eq:scalings_rs} with $\bar r = r-1$ such that $\gamma = \bar{\gamma}\,N^{-r}$, the matrix elements $a,\,b,\,d$ \eqref{eq:a_b_d_definition} appearing in eigenvalue expression~\eqref{eq:lpm-exact} become
\begin{equation}\label{app:eq:abd}
		\begin{aligned}
			a &= -\Bigl(1+\bar\gamma N^{-r}+i\,\bar\kappa N^{-s}\Bigr),\\
			b &= -\,\bar\gamma\,N^{\tfrac{1}{2}-r}\,\sqrt{1-\tfrac{1}{N}\,},\\
			d &= -\,\bar\gamma\,\Bigl(N^{1-r}-N^{-r}\Bigr).
		\end{aligned}
\end{equation}
This permits a direct assessment of the scaling of the imaginary parts of the eigenvalues $\lambda_{f,s}$ \eqref{eq:lpm-exact} in the asymptotic limit $N\rightarrow\infty$, thereby providing the time scaling $\tau=\Theta( |\Im\lambda_s|^{-1})$ via Eq.~\eqref{eq:Tdecaydef}. The six scaling regimes are summarized below.
	\begin{enumerate}[leftmargin=*]
		\item \textbf{Critical ($r=1,\ s=1/2$).} Set $\bar\gamma=1$. Here, at the exceptional point, $|a-d|\sim|b|\sim N^{-1/2}$ and both eigenvalues
		coalesce. Inserting this into Eq.~\eqref{eq:lpm-exact} and performing a Taylor expansion gives, to leading order in $N$,
		\begin{equation}\label{app:eq:lambda_pm_EP}
			\lambda_\pm = -1 \pm \frac{1}{\sqrt{N}}\sqrt{1-\frac{\bar\kappa^2}{4}} 
			- i\,\frac{\bar\kappa}{2\sqrt{N}}
			+ \mathcal{O}\bigl(N^{-1}\bigr).\hspace{-0.2cm}
		\end{equation} 
		This result yields $|\Im\lambda_\pm|=\Theta( N^{-1/2})$, giving the Grover-like scaling
		\begin{equation}
			\tau =\Theta( N^{\alpha})\,, \qquad \alpha = 1/2\,.
		\end{equation}
		Note that this scaling is very sensitive to $\bar \gamma$. To elaborate, suppose $\bar\gamma\neq1$. Now, to leading order in $N$ in the real and imaginary parts, the expansions for the two eigenvalues are		
		\begin{align}\label{eq:app:lambda_s_crit}
			\lambda_s =& -\bar{\gamma}-\frac{\bar{\gamma}}{(\bar{\gamma}-1) N} - i\,\frac{\bar{\gamma}^2 \bar{\kappa}}{(\bar{\gamma}-1)^2 N^{3/2}}\,,\\
			\lambda_f =& -1+\frac{\bar{\gamma}}{(\bar{\gamma}-1) N}-i\,\frac{\bar{\kappa}}{\sqrt{N}}\,,
		\end{align}
		where $\lambda_s=\lambda_+,\,\lambda_f=\lambda_-$ if $\bar\gamma<1$ and $\lambda_s=\lambda_-,\,\lambda_f=\lambda_+$ if $\bar\gamma>1$. The smaller of $|\Im\lambda_{\pm}|$ determines the slow mode $\lambda_s$ \eqref{eq:app:lambda_s_crit} that controls
		the long-time decay of $P(t)$ and hence the scaling exponent $\alpha$. Evidently, $\min\{|\mathrm{Im}\,\lambda_+|,|\mathrm{Im}\,\lambda_-|\}=|\Im\lambda_s|\sim N^{-3/2}$, and therefore choosing $\bar\gamma\neq1$ results in a slow-decaying mode with rate $N^{-3/2}$ that lives mostly in the symmetric subspace orthogonal to $\vert w\rangle$, preserving no-click amplitude. We obtain a \textit{worse-than-classical} scaling: $\tau= \Theta ( N^{\alpha}),\, \alpha = 3/2$. Note that this corresponds exactly to the scaling predicted by Regime B (cf.\ point 4.; Eq.~\eqref{app:eq:classB_scaling} with $r=1$, $s=1/2$), in which the exceptional point resides.
		
	\item \textbf{Regime A1 ($r=1,\ s>1/2$).} Set $\bar\gamma=1$. In this regime $|a-d|\sim N^{-s}\ll|b|\sim N^{-1/2}$ for $s\in(1/2,1]$ and $|a-d|\sim N^{-1}\ll|b|\sim N^{-1/2}$ when $s>1$, satisfying the expansions for ``strong-mixing'' in both cases. Therefore, eigenvalues take the form in Eq.~\eqref{eq:strong-mixing}, and both modes have an imaginary part $\Im a \simeq -\bar{\kappa}N^{-s}$ such that $|\Im\lambda_\pm| \sim N^{-s}$. The search time then asymptotically scales as 
		\begin{equation}
			\tau \sim N^{\alpha}\,, \qquad \alpha = s
		\end{equation}
		along the critical line $r=1$ with $\bar\gamma=1$. The time scaling is sublinear in the regime $s<1$.

        Now consider the case when $\bar\gamma\neq1$. To leading order we have $|a-d|\sim \mathcal O(1) \gg |b|$ (weak mixing). Consequently $\Im\lambda_s \simeq -\Im[b^2/(a-d)] = b^2 \Im(a-d)/|a-d|^2 \sim -N^{-1-s}$, and, recalling that $r=1$ here, the scaling behavior is governed by Regime B (cf.\ point 4.) with $\tau=\Theta( N^{s+1})$. This result indicates the sensitivity of the protocol to the choice of $\bar\gamma$, with search times worsening for $\bar\gamma\neq1$.
		
	\item \textbf{Regime A2 ($r=1,\ s<1/2$).} Again we start by taking $\bar\gamma=1$. Now $|a-d|\sim \bar\kappa\,N^{-s} \gg |b|\sim N^{-1/2}$, so the square root in \eqref{eq:lpm-exact} is dominated by the term $\sim |a-d|$. This is a case of ``weak-mixing" in the two-level non-Hermitian Hamiltonian. In accordance with expansion \eqref{eq:weak-mixing}, the slowest decay rate is given by $\mathrm{Im}\lambda_s \approx -b^2/(\bar\kappa N^{-s})\sim -N^{s-1}$. The dynamics quickly transfers weight to the slow mode $\lambda_s$, which controls the long-time behavior, and the search time is given by
		\begin{equation}\label{app:eq:classOverdamped_scaling}
        \tau =\Theta( N^{\alpha})\,, \qquad \alpha = 1-s\,.
		\end{equation}
		Note that the strongly damped eigenmode $\lambda_f$ decays at rate $\sim\kappa = \bar\kappa N^{-s}$, which yields a decay rate into target state $\vert w\rangle$ that is faster than the Grover scaling $\sim N^{1/2}$. However, the initial overlap of $\ket{s}$ with the $\lambda_f$ mode is small (of order $1/\sqrt{N}$): the system rapidly collapses into the long-lived slow mode and the long tail $\sim e^{-2|\mathrm{Im}\,\lambda_s|t}$ sets the search time $\tau$. For $s<1/2$, the exponent $\alpha=1-s > 1/2$ yields sublinear scaling. 

    Now suppose $\bar\gamma\neq1$. Based on earlier results (see points 1.\ and 2.), we anticipate the scaling to be sensitive to the choice of parameter $\bar\gamma$ along the critical line. Here we have $|a-d|\sim \mathcal{O}(1)$, so we still obtain weak mixing. However, $\Im (a-d)^{-1} = -\Im(a-d)/|a-d|^2\sim N^{-s}$, leading to $\Im \lambda_s \simeq -b^2\Im(a-d)^{-1}\sim -N^{-1-s}$. It follows directly that the search time is $\tau = \Theta(N^{s+1})$, and we recover the Regime B scaling; see point 4., Eq.~\eqref{app:eq:classB_scaling}. In terms of time complexity, this indicates a worsening in the algorithm performance compared to result~\eqref{app:eq:classOverdamped_scaling}, and it is best to fix $\bar\gamma=1$.

    \item \textbf{Regime B ($r>1,\ s>0$).}
	In this regime, asymptotically $b\sim N^{1/2-r}\to0$ and $a-d=\mathcal{O}(1)$. This corresponds to the condition of ``weak-mixing" $|a-d|\gg |b|$, and we perform perturbation theory according to Eq.~\eqref{eq:weak-mixing} to obtain the scaling of the slower mode: $\Im\lambda_s \approx - |b|^2 \Im \Bigl(\frac{1}{a-d}\Bigr) \sim -|b|^2\,\kappa \sim -N^{\,1-2r-s}$. It follows that the search time scales as
	\begin{equation}\label{app:eq:classB_scaling}
		\tau =\Theta( N^{\alpha})\,, \quad \alpha = 2r + s - 1 = 2\bar r +s +1\,.
	\end{equation}

    \item \textbf{Regime C ($r>s+1,\ s<0$).} Here $|a-d|\sim \bar\kappa\,N^{-s} \gg |b|\sim N^{1/2-r}$, which is satisfied when imposing the constraint $r>s+1\Rightarrow r>s+1/2$. Strong monitoring accelerates the decay of the term $e^{2\Im \lambda_f t}$ and the second-order contribution of $\Im \lambda_s$ dominates, setting the timescale. Using the same perturbative expansion \eqref{eq:weak-mixing}, $\Im\lambda_s \sim -|b|^2/\kappa \sim -N^{1-2r+s}$, and we find
	\begin{equation}
		\tau =\Theta( N^{\alpha})\,, \quad \alpha = 2r - s - 1=2\bar r -s +1\,.
	\end{equation}

    \item \textbf{Regime D ($r<1,\ s>r-1$).} Now $|a-d|\sim N^{1-r}\gg|b|\sim N^{1/2-r}$ for $N\rightarrow\infty$, so mixing is again weak. As a result, the eigenvalues are nondegenerate/``decoupled'' in the sense that one eigenvalue is connected continuously to $d$ and the other to $a$ as $b\to0$, and the relevant expansion is provided in Eq.~\eqref{eq:weak-mixing}. The mode with the largest overlap with the initial state, $\lambda_s$, determines the search time. Noting that asymptotically $\Im\Bigl(\frac{1}{a-d}\Bigr) = -\tfrac{\Im(a-d)}{|a-d|^2} \sim \tfrac{N^{-s}}{N^{2(1-r)}} =  N^{-2 + 2r - s}$, we find $\Im \lambda_s \approx -b^2\,\Im\Bigl(\frac{1}{a-d}\Bigr) \sim -N^{1-2r} N^{-2 + 2r - s} =- N^{-1 - s}$. To leading order in $N$, the timescale of the no-click probability decay yields
	\begin{equation}
		\tau =\Theta( N^{\alpha})\,, \qquad \alpha = s + 1\,.
	\end{equation}
\end{enumerate}

\begin{figure}[t]
    \centering
    \includegraphics[width=\linewidth]{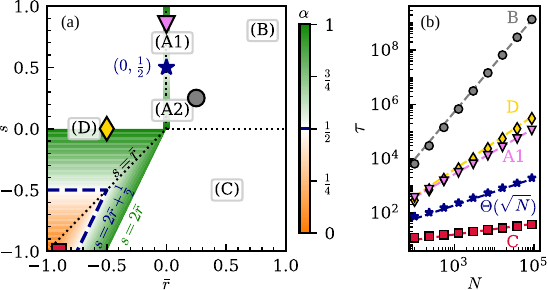}
    \caption{(a) Phase diagram of the asymptotic search time exponent $\alpha$ \textit{vs} exponents $\bar r\equiv r-1$ and $s$, with the color scale indicating the value of $\alpha$. The blue dashed line and star denote, respectively, the contour and point in the plane where $\tau=\tau_G$, i.e.\ $\alpha=1/2$. Unshaded regions have exponent $\alpha>1$.
Dotted lines separate different scaling regimes, labeled (B), (C), and (D), with scaling exponents summarized in Table~I of the main text. These lines represent continuous transitions in $\alpha$, except for the critical line at $\bar r=0,\,s\geq0$. (b) Search time $\tau$ as a function of system size $N$. Numeric data is shown for five points in parameter space; see the corresponding symbols in the $\bar r$-$s$ plane in (a) and Ref.~\cite{data_non_hermitian_search} for exact values of $\bar r$ and $s$. Dashed lines show the analytic scaling for each choice of $(\bar r,s)$ based on the results in Table~I of the main text.}
    \label{fig:phase_diagram_SM}
\end{figure}

\begin{figure*}[!htpb]
    \centering
    \includegraphics[width=\linewidth]{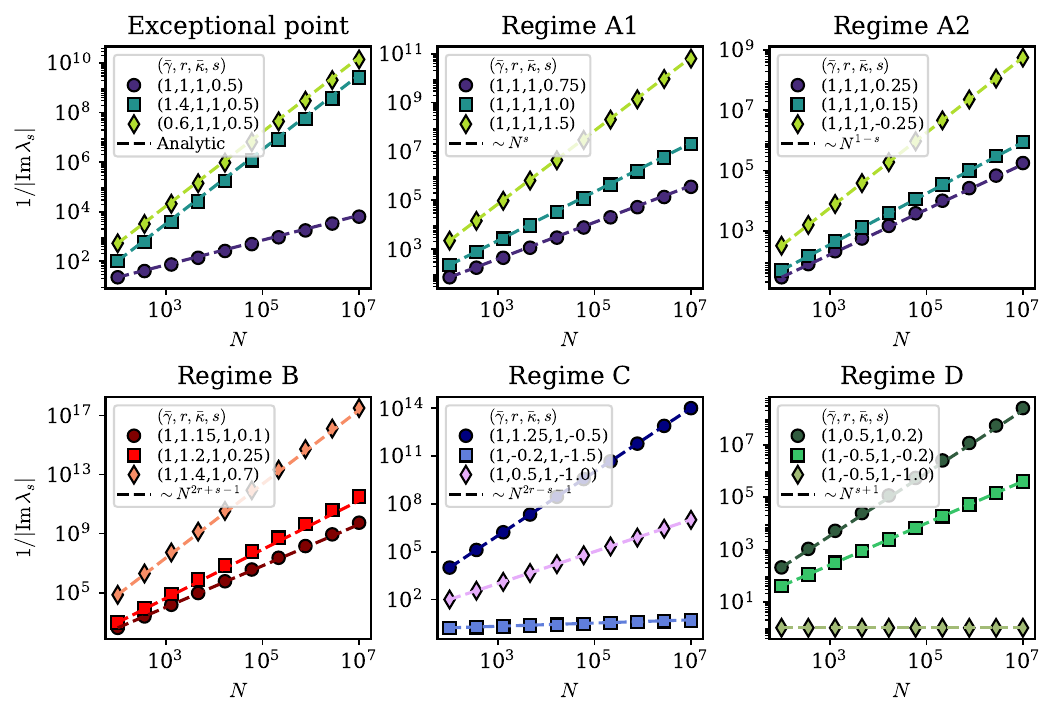}
    \caption{Comparison between the analytically predicted asymptotic scaling exponent $\alpha$---defined through the slow-mode behavior $|\Im \lambda_s|^{-1} \sim N^{\alpha}$---and the exact numerical eigenvalues 
	obtained by diagonalizing the effective Hamiltonian 
	$\hat{H}_{\mathrm{eff}}^{(s)}$ as a function of the system size $N$. The six panels correspond to the distinct scaling regimes: the top row shows the three cases along the critical line, while the bottom row presents the three scaling regimes~(B)--(D). Numerical results are shown as data points, with the corresponding analytic predictions represented as dashed lines in matching colors. Parameter values for $\gamma=\bar{\gamma} N^{-r}$ and $\kappa=\bar{\kappa} N^{-s}$ are provided in the legends of the respective panels.}
    \label{fig:Im_lambda_s_scaling}
\end{figure*}

\begin{figure*}[!htpb]
    \centering
    \includegraphics[width=\linewidth]{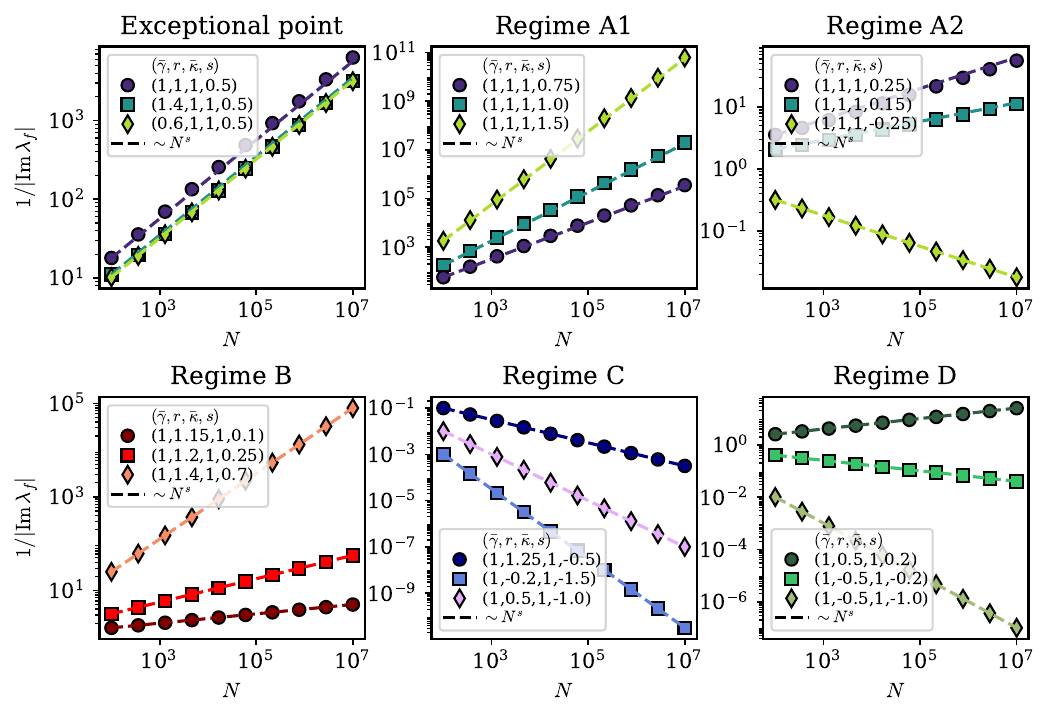}
    \caption{Comparison between the analytically predicted asymptotic scaling of the fast eigenmode $|\Im \lambda_f|^{-1} \sim N^{s}$ and the exact numerical eigenvalues obtained by diagonalizing the effective Hamiltonian $\hat{H}_{\mathrm{eff}}^{(s)}$ as a function of the system size $N$. Unlike the slow mode, the fast mode exhibits the same scaling behavior in all six scaling regimes. As in Fig.~\ref{fig:Im_lambda_s_scaling}, numerical results are shown as solid circles, with the corresponding analytic predictions represented as dashed lines in matching colors. Parameter values for $\gamma=\bar{\gamma} N^{-r}$ and $\kappa=\bar{\kappa} N^{-s}$ are provided in the legends of the respective panels.}
    \label{fig:Im_lambda_f_scaling}
\end{figure*}

\begin{figure*}[!htpb]
    \centering
    \includegraphics[width=\linewidth]{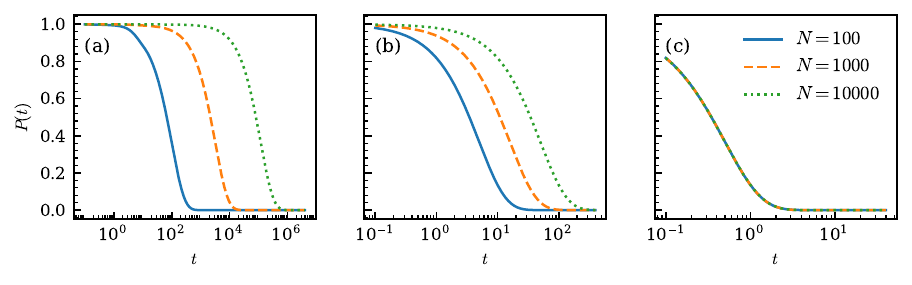}
    \caption{No-click probability decay dynamics, $P(t)$ \textit{vs} time $t$, in three different parameter regimes with hopping rate $\gamma = N^{-r}$ and monitoring rate  $\kappa = N^{-s}$ for three representative values of the system size $N$ (see legend).  Panels: (a) We fix $r = 1.1, s = 0.25$, placing us in scaling regime (B). Here, the convergence time $\tau$ of the no-click probability to some small threshold value increases as $\Theta(N^\alpha)$, $\alpha>1$, with system size, making the scaling inferior to classical search algorithms. (b) Setting $r = -1.0, s = -0.5$, we are in regime (D) where we recover Grover scaling of $\alpha = 1/2$. (c) For $r = -1.0, s = -1.0$ we are in the boundary of regime (C) and (D). Here, $\alpha = 0$, that is the search time is better than Grover scaling and independent of $N$. The scaling of the search time with $N$ for a threshold value $\varepsilon  = 0.001$ is demonstrated in Fig.~(2)(b) of the main text.}
    \label{fig:SP_three_N} 
\end{figure*}

These results for the search time are summarized in Table I of the main text. In Fig.~\ref{fig:Im_lambda_s_scaling} we verify the analytic scaling exponent $\alpha$ by comparing with the exact numeric eigenvalues obtained by diagonalizing the Hamiltonian $\hat H_{\rm eff}^{(s)}$. The extracted scaling $|\Im\lambda_s|^{-1}\sim N^\alpha$ agrees well with the numerical data. The scaling of the so-called \textit{fast eigenmode} $\lambda_f$ does not influence the search time in the absence of resetting; however, it will be relevant later when we introduce the resetting protocol. Independent of the exponent values $r$ and $s$, this second eigenvalue scales as $|\Im\lambda_f|^{-1}\sim N^s$; see Fig.~\ref{fig:Im_lambda_f_scaling}. We also confirmed in Fig.~\ref{fig:phase_diagram_SM}(b) (refer to symbols in phase diagram \ref{fig:phase_diagram_SM}(a) for parameter values and regimes in which the various data points lie) that the exact numeric value for the time taken for $P(t)$ to reach a threshold value $\varepsilon$ (we use $\varepsilon = 0.001$ in the numerics) indeed scales as $\tau=\Theta(N^\alpha)$. Finally, to illustrate the no-click probability dynamics, Fig.~\ref{fig:SP_three_N} shows $P(t)$ as a function of time for three different system sizes $N$.

\subsection{Overlaps}\label{app:ss:overlaps}
We compute the asymptotic scaling of the overlaps in the no-click probability $P(t)$. In accordance with Eqs.~\eqref{eq:surv_prob_expand} and \eqref{eq:surv_prob_simplified_appendix}, these are defined as $O_j = \langle\lambda_j|\lambda_j\rangle\langle s|\bar\lambda_j\rangle\langle\bar\lambda_j|s\rangle$, $j\in\{+,-\}$, and $O_\pm = \langle\lambda_-|\lambda_+\rangle \langle s|\bar\lambda_-\rangle\langle\bar\lambda_+|s\rangle$. The equal superposition state is $|s\rangle=(1/\sqrt{N},\sqrt{1-1/N})^\intercal$. 
The asymptotic scaling will depend strongly on the behavior of the left and right eigenvectors, $\langle \bar\lambda_\pm\vert$ and $\vert\lambda_\pm\rangle$, of the effective Hamiltonian $\hat H_{\mathrm{eff}}^s$. Define the (unnormalized) right eigenvectors as $\vert \lambda_\pm' \rangle= (v_\pm,\,1)^\intercal$, with
\begin{equation}\label{eq:vpm_exact}
	v_\pm = \frac{\lambda_\pm -d}{b}\,. 
\end{equation}
It follows that the left eigenvectors are $\langle \bar\lambda_\pm'\vert= (v_\pm,\,1)$. We write the inner products appearing in the overlaps $O_j$ in terms of the components $v_\pm$ \eqref{eq:vpm_exact} as
\begin{align}\label{eq:inner_product_Oj_1}
    \langle\lambda_j|\lambda_j\rangle
		&= \frac{\langle\lambda_j'|\lambda_j'\rangle}{|\mathcal N_j|}
		= \frac{1+|v_j|^2}{|v_j^2+1|},\\
	\langle s|\bar\lambda_j\rangle\langle\bar\lambda_j|s\rangle
		&= \frac{|\langle s|\bar\lambda_j'\rangle|^2}{|\mathcal N_j|} 
		= \frac{\big|\sqrt{1-\tfrac{1}{N}}+\tfrac{v_j^*}{\sqrt{N}}\big|^2}{|v_j^2+1|},\label{eq:inner_product_Oj_2}
\end{align}
where $\mathcal N_j=\langle\bar\lambda_j'|\lambda_j'\rangle = v_j^2+1$ is the normalization factor, see Eq.~\eqref{eq:eigenvector_normalization}, and $\langle s|\bar\lambda_j'\rangle = \sqrt{1-1/N} + v_j^*/\sqrt{N}$ follows directly from the definitions of $|s\rangle$ and $|\bar \lambda_j'\rangle$. 

Multiplying the two expressions \eqref{eq:inner_product_Oj_1} and \eqref{eq:inner_product_Oj_2} yields the convenient (and real) form
\begin{equation}\label{eq:Oj_analytic}
	O_j = \frac{(1+|v_j|^2)\,\big|\sqrt{1-\tfrac{1}{N}}+\tfrac{v_j}{\sqrt{N}}\big|^2}{|1+v_j^2|^2}\,, \quad j\in\{+,-\}.
\end{equation}
Since $v_\pm$ are in general complex, it will be convenient to introduce the polar decomposition of the eigenvector components
\begin{equation*}
    v_j = r_j e^{i\phi_j},\qquad r_j\equiv |v_j|,\;\; \phi_j\equiv\arg(v_j).
\end{equation*}
In this representation, the overlaps are given by
\begin{equation}\label{eq:Oj_polar}
    O_j = \frac{(1+r_j^2)\Big(1-\tfrac{1}{N}+\tfrac{r_j^2}{N}
			+ \tfrac{2r_j}{\sqrt{N}}\sqrt{1-\tfrac{1}{N}}\cos\phi_j\Big)}{1 + r_j^4 + 2 r_j^2\cos(2\phi_j)}.
\end{equation}

\begin{figure*}
    \centering
    \includegraphics[width=\linewidth]{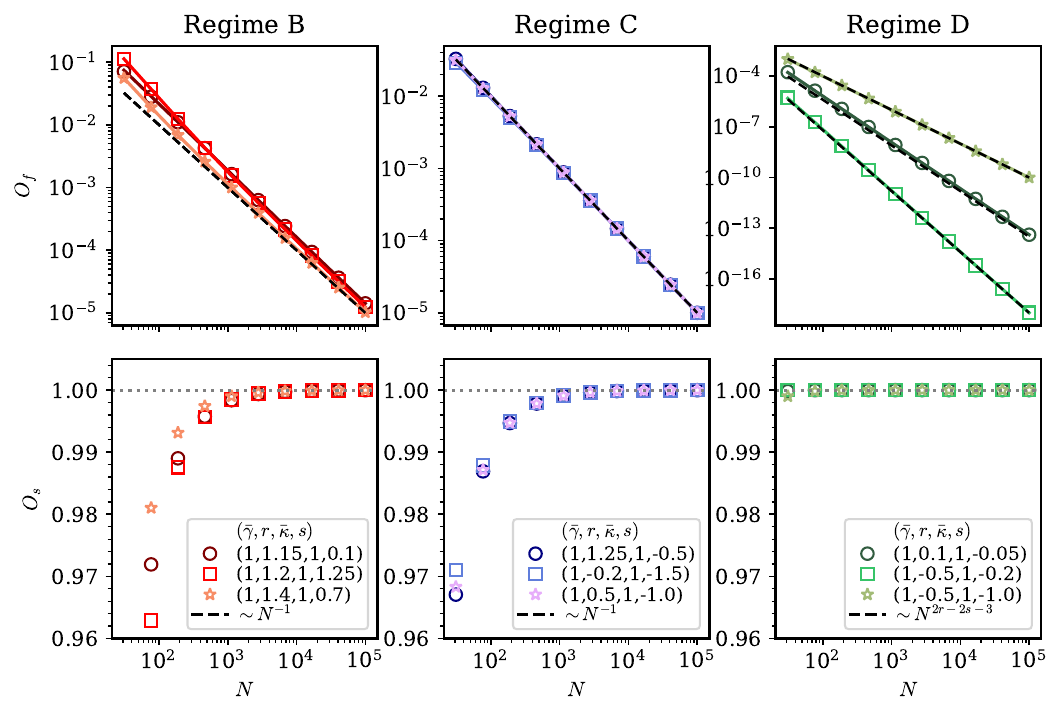}
    \caption{Scaling of the fast–mode (top row) and slow-mode (bottom row) overlaps, $O_f$ and $O_s$, respectively, with system size $N$ across the three scaling regimes. Each column corresponds to a single scaling regime: Regime~B (left), Regime~C (center), and Regime~D (right). Symbols denote exact numerical results obtained by diagonalizing the effective non-Hermitian Hamiltonian $\hat H_{\mathrm{eff}}^{(s)}$ and computing the inner products $O_j = \langle\lambda_j|\lambda_j\rangle\langle s|\bar\lambda_j\rangle\langle\bar\lambda_j|s\rangle$, $j\in\{f,s\}$, while dashed black lines show the analytically predicted asymptotic behavior~\eqref{app:eq:asymptotic_Of_Os}. Colored solid lines in the top row demonstrate the overlaps computed via the analytic expression \eqref{eq:Oj_analytic}. Different markers and colors distinguish parameter sets $(\bar{\gamma}, r, \bar{\kappa}, s)$ with $\gamma = \bar{\gamma} N^{-r}$ and $\kappa = \bar{\kappa} N^{-s}$, as listed in the legends. $O_f$ decreases with $N$, while $O_s$ approaches a constant in the thermodynamic limit, consistent with result~\eqref{app:eq:asymptotic_Of_Os}.}
    \label{fig:overlaps}
\end{figure*}

Disregarding the critical cases, which are discussed in more detail in Sec.~\ref{app:6}, we focus on regimes (B)--(D) in which there is ``weak-mixing" $|a-d|\gg|b|$. For convenience, we again adopt the subscript notation $f,s$, as opposed to $\pm$, to distinguish the two overlaps based on the corresponding decay rate of the exponential function. Accordingly, we have
\begin{equation}
    O_s = \begin{cases}
        O_+\,, \quad & \mathrm{if}\ |\Im\lambda_+|\leq |\Im\lambda_-|\,,\\
        O_-\,, &\mathrm{otherwise}\,,
        \end{cases}
\end{equation}
for the prefactor of the exponential function that decays with the ``slow'' eigenvalue, while the overlap for the fast term is
\begin{equation}
    O_f = \begin{cases}
        O_+\,, \quad & \mathrm{if}\ |\Im\lambda_+|\geq |\Im\lambda_-|\,,\\
        O_-\,, &\mathrm{otherwise}\,.
    \end{cases}
\end{equation}
Treating $|b|$ as a small perturbation as before, the eigenvalues to $\mathcal{O}\left(\tfrac{b^4}{(a-d)^3}\right)$ are $\lambda_{f} = a + \tfrac{b^2}{a-d}$ and $\lambda_{s} = d - \tfrac{b^2}{a-d}$; see Eq.~\eqref{eq:weak-mixing}. From this we obtain the leading-order forms for the vector components:

\begin{equation}
    \label{eq:vpm_approx}
		v_f\simeq\frac{a-d}{b}+\frac{b}{a-d}\,,\qquad
		v_s\simeq -\frac{b}{a-d}\,,
\end{equation}
Using Eq.~\eqref{eq:vpm_approx}, we determine the scaling behavior of the overlaps $O_f$ and $O_s$ \eqref{eq:Oj_polar}, corresponding to the prefactors of the fast and slow exponential decay channels, in the three regimes: 
\begin{itemize}[leftmargin=0.7cm]
    \item[(B)] $r>1,\; s>0$: This case is more straightforward than the former case, and we work in the polar representation~\eqref{eq:Oj_polar}. Note that $|a-d|\sim \mathcal{O}(1)$ and $|b|\sim N^{1/2-r}$. With this scaling, we can infer that $1+r_f^2 \sim r_f^2$, $1 + r_f^2/N + 2r_f/\sqrt{N}\cos\phi_f \sim r_f^2/N$ and $1+r_f^4+2r_f^2\cos(2\phi_f)\sim r_f^4$, which leads to the asymptotic results for the fast mode overlap $O_f \sim r_f^2\cdot (r_f^2/N)/r_f^4 \sim N^{-1}$. Since $r_s\ll1$, it trivially follows that $O_s\sim\mathcal O(1)$.
    
    \item[(C)] $s<0,\; r>s+1$: As before, $|a-d|\sim N^{-s}$ and $|b|\sim N^{1/2-r}$. For valid exponent values $r$ and $s$, one can show that $r_f\sim N^{r-1/2-s}\gg1$ and $r_s\sim N^{1/2-r+s} \ll1$. Inserting the asymptotics into Eq.~\eqref{eq:Oj_polar} and keeping terms to leading order in $N$ yields $O_f \sim N^{-1}$ and $O_s\sim\mathcal O(1)$. 

	\item[(D)] $r<1,\; s>r-1$: The asymptotic scaling of the overlap $O_f$ in Regime (D) is particularly subtle. Naively one might take $|a-d|\sim N^{1-r}$ and approximate $|b|\sim N^{1/2-r}$ in the asymptotic $N\to\infty$ limit, implying that $r_f\equiv |v_f| \sim N^{1/2}$ and $r_s\equiv |v_s| \sim N^{-1/2}$. For the fast-decaying mode we then find $1+r_f^2 \sim r_f^2 \sim N$,  $1 + r_f^2/N + 2r_f/\sqrt{N}\cos\phi_f \sim 1 + \mathcal O(1) \sim \mathcal O(1)$, and $1+r_f^4+2r_f^2\cos(2\phi_f)\sim r_f^4\sim N^2$. It follows that $O_f \sim \frac{N\cdot \mathcal O(1)}{N^2} \sim N^{-1}$. In the case of $O_s$, we compute $1+r_s^2 \sim \mathcal O(1)$, $1 + r_s^2/N + 2r_s/\sqrt{N}\cos\phi_s \sim\mathcal O(1)$, and $1 + r_s^4 + 2 r_s^2\cos(2\phi_s)\sim \mathcal O(1)$. Consequently, the overlap is $O_s\sim\mathcal O(1)$. However, the naive approximation $|b|\sim N^{1/2-r}$ actually yields only an upper bound on the fast overlap $O_f\lesssim N^{-1}$. This is due to the sub-leading term in the square root $\sqrt{1-1/N}$, see Eq.~\eqref{app:eq:abd}, which is essential to obtain a cancellation of the terms of order $1/N$. 

    More rigorously, let us instead approximate $\sqrt{1-1/N}\simeq 1-\tfrac{1}{2N}$ in $b$~\eqref{app:eq:abd} and work with the analytic expression~\eqref{eq:Oj_analytic}. The vector component $v_f$ depends on $a$, $b$ and $d$, now defined as $a=-(1+\bar\gamma N^{-r}+i\,\bar\kappa N^{-s})$, $b\simeq -\bar\gamma N^{\tfrac{1}{2}-r}(1-\frac{1}{2N})$, and $d=-\bar\gamma(N^{1-r}-N^{-r})$, respectively.  Define the dominant real scale of $a-d$ by $A_0 \equiv \bar\gamma N^{1-r}$. It follows that sub-leading terms are $\xi \equiv (a-d)-A_0 = -1 -2\bar\gamma N^{-r} - i\bar\kappa N^{-s}$. The difference between $a$ and $d$ is then compactly written as $a-d = A_0 + \xi$, with the ratio of the two terms $\xi/A_0 = -\dfrac{1}{\bar\gamma}N^{r-1} -2N^{-1} - i\dfrac{\bar\kappa}{\bar\gamma}N^{r-s-1}$. In this notation $b\simeq -A_0 N^{-1/2}(1-\frac{1}{2N})$. We proceed by explicitly computing the two terms of the vector component $v_f$ \eqref{eq:vpm_approx}, which directly determine the overlap $O_f$ via Eq.~\eqref{eq:Oj_analytic}:
    \begin{equation*}
        \frac{a-d}{b}
	= \frac{A_0+\xi}{-A_0 N^{-1/2}(1-\tfrac{1}{2N})}
	= -N^{1/2}\frac{1+\varepsilon}{1-\tfrac{1}{2N}},
    \end{equation*}
    \begin{equation*}
        \frac{b}{a-d} = \frac{-A_0 N^{-1/2}(1-\tfrac{1}{2N})}{A_0(1+\varepsilon)}
	= -N^{-1/2}\frac{1-\tfrac{1}{2N}}{1+\varepsilon},
    \end{equation*}
    with $\varepsilon\equiv \xi/A_0$. In the first equation, we now use the expansion $\dfrac{1}{1-\frac{1}{2N}}=1+\dfrac{1}{2N}+\mathcal O(N^{-2})$, substitute the exact expression for the ratio $\varepsilon = \xi/A_0$, and simplify to leading order, keeping all constant ($r$- and $s$-independent) terms up to $\mathcal{O}(N^{-3/2})$. This yields $(a-d)/b\sim -N^{\tfrac{1}{2}} + \frac{1}{\bar\gamma}N^{r-\tfrac12} + \frac{3}{2}N^{-\tfrac12}
	+ i\frac{\bar\kappa}{\bar\gamma}N^{r-s-\tfrac12}$. Focusing on the second equation, $|\varepsilon|\ll1$ since $\xi/A_0$ is a sub-leading term in $a-d$, and we therefore perform perturbation theory. Expanding $1/(1+\varepsilon)=1-\varepsilon+\varepsilon^2+\cdots$, substituting the explicit expression for $\varepsilon$ and working to the same order as the first equation, we obtain $b/(a-d)\sim -N^{-\tfrac{1}{2}}$. We combine the two asymptotic results, see Eq.~\eqref{eq:vpm_approx}, to give $v_f/\sqrt{N}\sim -1 + \frac{1}{\bar\gamma}N^{r-1} + \frac{1}{2}N^{-1}
	+ i\frac{\bar\kappa}{\bar\gamma}N^{r-s-1}$. Finally, to obtain the asymptotic scaling of the overlap $O_f$~\eqref{eq:Oj_analytic}, we expand $\sqrt{1-1/N}\simeq 1-\tfrac{1}{2N}$ as before and compute $\frac{(1+|v_f|^2)}{|1+v_f^2|^2}\sim N^{-1}$ and $\bigg|\sqrt{1-\frac{1}{N}}+\frac{v_f}{\sqrt N}\bigg|^2
		\sim \Big|\frac{1}{\bar\gamma}N^{\,r-1} + i\frac{\bar\kappa}{\bar\gamma}N^{\,r-1-s}\Big|^2$. Up to sub-leading corrections, it follows from Eq.~\eqref{eq:Oj_analytic} that 
        \begin{equation}\label{app:eq:Of_classA}
            O_f\sim \begin{cases}
		    N^{2r-3},\quad &\text{if } s>0,\\
            N^{2r-2s-3},& \text{if } r-1<s<0.
		\end{cases}
        \end{equation}
    This demonstrates that $N^{-1}$ is an upper bound on the overlap $O_f$ in this case, but a more rigorous asymptotic analysis yields the leading order behavior in Eq.~\eqref{app:eq:Of_classA}. The scaling of the slow overlap is $O_s\sim\mathcal{O}(1)$. These results are confirmed by exact numeric results; see Fig.~\ref{fig:overlaps}.

\end{itemize}
To summarize, in the perturbative regime, $|b|\ll|a-d|$, one finds generically
\begin{align} \label{app:eq:asymptotic_Of_Os}
O_f&\xrightarrow{N\to\infty} \begin{cases}
    \mathcal{O}(N^{2r-3}),\quad\quad \text{ if } s>0, \quad & [\text{Regime D}]\\ 
    \mathcal{O}(N^{2r-2s-3}), \quad \text{if } s<0,\quad & [\text{Regime D}]\\ 
    \mathcal{O}(N^{-1}),&[\text{Regime B, C}]
\end{cases}\nonumber\\
O_s&\xrightarrow{N\to\infty}\mathcal{O}(1). \qquad [\text{All regimes}]
\end{align}
These results for the overlaps confirm that the weight $O_s$ of the \textit{slow term} in the no-click probability~\eqref{eq:surv_prob_simplified_appendix}, i.e., the term with the slowest rate of exponential decay,  is always large relative to the weight $O_f$ of the \textit{fast term}. In fact, in the limit $N\to\infty$, $O_f\to0$ and $O_s\to1$ in the entire $(r,s)$ parameter space, excluding the critical line $r=1,\,s\geq0$. The latter is a special case in which $O_f\approx O_s\sim\mathcal{O}(1)$, and is treated in detail in Sec.~\ref{app:6}. Numeric results confirm the analytic scaling; see Fig.~\ref{fig:overlaps}.

To close this section, we briefly turn our attention to the interference term. We introduce a bound on the overlap of this term, $O_\pm = \langle\lambda_-|\lambda_+\rangle \langle s|\bar\lambda_-\rangle\langle\bar\lambda_+|s\rangle$. This will confirm that, in the asymptotic limit, the interference term is negligible compared to the ``slow term'' in $P(t)$ \eqref{eq:surv_prob_simplified_appendix}, which decays on a timescale governed by the slow eigenvalue, $\Im\lambda_s$. In many cases, it is therefore permitted to drop this term. Letting $\eta \in [0,1]$ be the cosine of the principal angle between the right eigenvectors
\begin{equation}
	\eta \coloneqq \frac{\bigl| \langle \lambda_- \,|\, \lambda_+ \rangle \bigr|}
			{\sqrt{\langle \lambda_+|\lambda_+\rangle\,\langle \lambda_-|\lambda_-\rangle}}\,,
\end{equation}
we find the bound
\begin{equation}
	|O_{\pm}| = \eta\, \sqrt{O_+\,O_-}\le \sqrt{O_+\,O_-}\,.
	\label{eq:Opm_bound}
\end{equation}
This bound is determined by the Cauchy–Schwarz inequality 

$\bigl| \langle \lambda_-|\lambda_+\rangle \bigr|\le \|\lambda_-\|\,\|\lambda_+\|$, with the vector norm $\|\lambda_j\| \equiv \sqrt{\langle \lambda_j\vert\lambda_j\rangle}$, such that
\begin{align*}
	|O_{\pm}|
	&= |\langle \bar\lambda_+|s\rangle|\, |\langle s|\bar\lambda_-\rangle|\,
			| \langle \lambda_-|\lambda_+\rangle |\\
	&=\sqrt{ |\langle s|\bar\lambda_+\rangle|^2 \,\|\lambda_+\|^2}\;\sqrt{ |\langle s|\bar\lambda_-\rangle|^2 \,\|\lambda_-\|^2}\; \frac{|\langle \lambda_-|\lambda_+\rangle|}{\|\lambda_-\|\,\|\lambda_+\|}\\
	&= \eta\,\sqrt{O_+\,O_-}\\
	&\le
			|\langle \bar\lambda_+|s\rangle|\,|\langle s|\bar\lambda_-\rangle|\,
			\|\lambda_-\|\,\|\lambda_+\|\\
	&= \sqrt{O_+\,O_-}\,.
\end{align*}

Let us unpack the implications of this result~\eqref{eq:Opm_bound}. Consider parameter values away from the exceptional point, i.e., choices for $\gamma$, $\kappa$ that lead to scaling regimes (B)--(D); see itemized list in Sec.~\ref{app:ss:asymptotics_evs} and \ref{app:ss:overlaps}. Immediately we note that $|O_\pm|\sim \sqrt{O_f}$ (since $O_s\sim\mathcal O(1)$ according to Eq.~\eqref{app:eq:asymptotic_Of_Os}), which tends to zero in the asymptotic limit $N\rightarrow\infty$. Refer to Eqs.~\eqref{app:eq:asymptotic_Of_Os} and \eqref{eq:Opm_bound}. To make this more concrete in the context of the no-click probability, let us consider the complete interference term, see Eq.~\eqref{eq:surv_prob_simplified_appendix}. Using $|{\rm Re}(z)|\le |z|$, $|e^{-i z t}|=e^{\,\mathrm{Im}(z)t}$ and result \eqref{eq:Opm_bound}, the interference term is bounded from above by
\begin{equation}
    \Bigl|2\,\mathrm{Re}\bigl[ O_{\pm}\,e^{-i(\lambda_+ - \lambda_-^*)t}\bigr]\Bigr|\le
				2e^{(\mathrm{Im}\lambda_+ + \mathrm{Im}\lambda_-)t}\sqrt{O_+\,O_-}\,.
\end{equation}
Taking the ratio of the slow and interference terms, we obtain
\begin{align}
    \frac{\rm interference}{\rm slow\; term} =\;& \frac{
		\Bigl|2\,\mathrm{Re}\bigl[ O_{\pm}\,e^{-i(\lambda_+ - \lambda_-^*)t}\bigr]\Bigr|}{O_{s}\,e^{2\Im\lambda_s t}}\nonumber\\
		\le &\; 2 \sqrt{\frac{O_{f}}{O_{s}}}\;
		e^{(\Im\lambda_f-\Im\lambda_s)t}.
\end{align}
By definition, $\Im\lambda_f<\Im\lambda_s$ away from the exceptional point, so the interference term (i) decays monotonically in $t$ and (ii) has an amplitude that tends to zero in the large-$N$ limit. Compared to the \textit{slow term}, the interference term can be seen as negligible, and the rate of the slow mode reliably represents the timescale of the search as claimed earlier; see Eq.~\eqref{eq:Tdecaydef}.

\section{Exceptional point: Eigenvalues, overlaps and time complexity bounds}\label{app:6}
Suppose we are now at the exceptional point $(\gamma,\kappa) = (\gamma^{\rm EP},\kappa^{\rm EP})$. The imaginary parts of the eigenvalues exhibit the same scaling with $N$; see point 1.\ of Sec.~\ref{app:ss:asymptotics_evs}. In addition, the overlaps $O_f,\; O_s$ are both $\mathcal{O}(1)$, as we elaborate below. Consequently, the interference term cannot be neglected. This case is interesting, both algebraically and physically, since it is the only point along $r =0$ where Grover-like time complexity is recovered for our dynamics. We proceed by (i) computing the overlaps, showing they are $\mathcal{O}(1)$, (ii) demonstrating that the interference term is not negligible, and (iii) computing upper and lower bounds on the search time, thereby confirming $\tau=\Theta(\sqrt N)$ at the exceptional point.

First, set $\gamma = 1/N$ and $\kappa = \bar{\kappa} / \sqrt{N}$, with $\bar{\kappa} \in [0,2)$. This satisfies the requirement for being at the exceptional point. Working in the large-$N$ limit and to $\mathcal{O}(N^{-3/2})$, the eigenvectors of the effective Hamiltonian take the form
\begin{equation}\label{app:eq:evec:EP}
    \ket{\lambda_\pm} \simeq \mathcal{N}_\pm  \begin{bmatrix}
        \mp \frac{\sqrt{4-\bar{\kappa}^2}}{2} + \frac{1}{\sqrt{N}}  + i \left( \frac{\bar{\kappa}}{2} \mp \frac{\bar{\kappa}}{\sqrt{(4-\bar{\kappa}^2) N}} \right)\\
        1
     \end{bmatrix},
\end{equation}
with the left eigenvectors following as $\ket{\bar\lambda_\pm}=\ket{\lambda_\pm}^*$ and $\mathcal N_\pm$ the normalization (see Sec.~\ref{app:2}). Using $\ket{s} = \big( \frac{1}{\sqrt{N}}, 1 - \frac{1}{2N} \big)^\intercal$ and expanding in terms of small parameter $1/N$, the overlaps $O_j = \langle\lambda_j|\lambda_j\rangle\langle s|\bar\lambda_j\rangle\langle\bar\lambda_j|s\rangle$, $j\in\{+,-\}$, are
\begin{align}\label{eq:EP_Op}
    O_{+}&= \frac{2}{4-\bar{\kappa}^2}+\frac{2 \left(\bar{\kappa}^2-2\right)}{\left(4-\bar{\kappa}^2\right)^{3/2}\sqrt{N}}+\frac{4 \left(3 \bar{\kappa}^2-4\right)}{\left(\bar{\kappa}^2-4\right)^3 N},\\
	O_{-} &= \frac{2}{4-\bar{\kappa}^2}-\frac{2 \left(\bar{\kappa}^2-2\right)}{\left(4-\bar{\kappa}^2\right)^{3/2}\sqrt{N}}+\frac{4 \left(3 \bar{\kappa}^2-4\right)}{\left(\bar{\kappa}^2-4\right)^3 N},
    \label{eq:EP_Om}
\end{align}
where we have dropped sub-leading terms of order $\mathcal{O}(N^{- 3/2})$ and higher. Since $\bar\kappa$ is an $\mathcal{O}(1)$ constant, we find that both overlaps are of order $1$ to leading order. Hence, using the \textit{slow} and \textit{fast} mode terminology introduced previously, we conclude that $O_s\approx O_f \sim \mathcal{O}(1)$ and both exponential terms in $P(t)$~\eqref{eq:surv_prob_simplified_appendix} contribute significantly to the no-click probability. 

The overlap of the interference term, $O_\pm = \langle\lambda_-|\lambda_+\rangle \langle s|\bar\lambda_-\rangle\langle\bar\lambda_+|s\rangle$, is similarly computed from the eigenvectors in Eq.~\eqref{app:eq:evec:EP}. The real and imaginary parts of $O_\pm$ to $\mathcal{O}(N^{- 3/2})$ are found to be
\begin{align}\label{eq:EP_Re_opm}
    \text{Re} \big[ O_{\pm} \big] &= \frac{\left(\bar{\kappa}^2-3\right) \bar{\kappa}^2}{2 \left(\bar{\kappa}^2-4\right)}  \\ &\phantom{=}+\frac{88 \bar{\kappa}^2+\left(\bar{\kappa}^2-6\right) \left(\bar{\kappa}^4-4 \bar{\kappa}^2+12\right) \bar{\kappa}^4-32}{2 \left(\bar{\kappa}^2-4\right)^3 N} ,\nonumber\\
	\text{Im} \big[ O_{\pm}\big] &= -\frac{\bar{\kappa} \left(\bar{\kappa}^2-1\right)}{2 \sqrt{4-\bar{\kappa}^2}}\\
    &\phantom{=}-\frac{\bar{\kappa} \left(\bar{\kappa}^8-8 \bar{\kappa}^6+22 \bar{\kappa}^4-40 \bar{\kappa}^2+32\right)}{2 \left(4-\bar{\kappa}^2\right)^{5/2} N},\nonumber
\end{align}
with $\Re O_{\pm}$ being the quantity of interest since it enters directly in the third term of $P(t)$~\eqref{eq:surv_prob_simplified_appendix}. Recalling that $\Im \lambda_\pm \sim - \frac{\bar\kappa}{2\sqrt{N}}$ and $-i(\lambda_+ - \lambda_-^*) \sim -\frac{ \bar{\kappa}}{\sqrt{N}}-i\sqrt{\frac{4-\bar{\kappa}^2}{N}}$, see point 1.\ in Sec.~\ref{app:ss:asymptotics_evs} and Eq.~\eqref{app:eq:lambda_pm_EP}, we insert the leading-order expressions for the eigenvalues and overlaps, cf.\ Eqs.\ \eqref{eq:EP_Op}-\eqref{eq:EP_Re_opm}, into the no-click probability~\eqref{eq:surv_prob_simplified_appendix} and simplify:
\begin{align}\label{app:eq:asymptotic_Pt_EP}
    P(t) =&\; e^{-\bar{\kappa}t/\sqrt{N}}\left(\frac{4}{4-\bar{\kappa}^2}\right) \nonumber\\
    &+  e^{-\bar{\kappa}t/\sqrt{N}} \left[ \cos\left(\sqrt{\frac{4-\bar{\kappa}^2}{N}}t\right)\frac{\left(\bar{\kappa}^2-3\right) \bar{\kappa}^2}{\left(\bar{\kappa}^2-4\right)} \right] \nonumber \\ 
	&+ e^{-\bar{\kappa}t/\sqrt{N}}  \left[ \sin\left(\sqrt{\frac{4-\bar{\kappa}^2}{N}}t\right)\frac{\bar{\kappa} \left(\bar{\kappa}^2-1\right)}{\sqrt{4-\bar{\kappa}^2}} \right].
\end{align}
The last two terms originate from the interference term, and evidently have a non-negligible contribution to $P(t)$, even for $N\to\infty$.

\begin{figure}
    \centering
    \includegraphics[width=\linewidth]{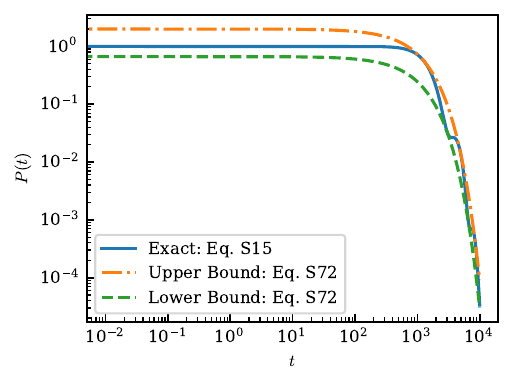}
    \caption{No-click probability $P(t)$ plotted against time with $(\gamma,\kappa)$ corresponding to the exceptional point. The orange dot-dashed (green dashed) line represents the upper (lower) bound to the no-click probability as written in Eq.~\ref{eq:bounds}. Other parameters are $N = 10^6, \gamma = 1/N, \kappa = 1/\sqrt{N}$ and for the numerical simulation $\rm{d}t = 0.01$.  }
    \label{fig:bound}
\end{figure}

The asymptotic expression for the no-click probability \eqref{app:eq:asymptotic_Pt_EP} can also be used to determine an upper and lower bound on the search time $\tau$. Express Eq.~\eqref{app:eq:asymptotic_Pt_EP} compactly as
\begin{equation}\label{eq:Scompact}
	P(t)=e^{-a t}\bigl(A+B\cos(\omega t)+C\sin(\omega t)\bigr),
\end{equation}
where
	\begin{align*}
		a&=\frac{\bar\kappa}{\sqrt{N}}, &  A&=\frac{4}{4-\bar\kappa^2},\\
		\omega&=\frac{\sqrt{4-\bar\kappa^2}}{\sqrt{N}}, & B&=\frac{(\bar\kappa^2-3)\bar\kappa^2}{\bar\kappa^2-4},\\
		&& C&=\frac{\bar\kappa(\bar\kappa^2-1)}{\sqrt{4-\bar\kappa^2}}.
	\end{align*}
Reformulating this in terms of a single cosine function, we have
\begin{equation}
    P(t)=e^{-a t}\bigl(A + D\cos(\omega t-\phi)\bigr)
\end{equation}
with $D=\sqrt{B^2+C^2}=\frac{2\bar\kappa}{4-\bar\kappa^2}$ and $\phi=\arctan(C,B)$. $P(t)$ cannot be inverted to yield a single closed-form solution for the time $t$, making it challenging to analytically extract the search time $\tau$. However, we can obtain upper and lower bounds on the computational time by using the oscillation envelope and thereby determine the asymptotic scaling of the search time.  Since $-1\le\cos(\cdot)\le1$ and $A,D>0$ for $\bar{\kappa}\in(0,2)$, the no-click probability is bounded above and below as
\begin{equation}\label{eq:bounds}
	e^{-a t}(A-D)\le P(t)\le e^{-a t}(A+D).
\end{equation}
Fig.~\ref{fig:bound} confirms that both upper and lower bounds constrain the no-click probability tightly, with the agreement becoming even better at long times. Thus, if $P(t)=c$ has a solution, with $c$ a constant interpreted as a chosen fixed target probability, a necessary condition is $A-D \le c\,e^{a t}\le A+D$, giving $\frac{1}{a}\ln\frac{A-D}{c}\ \le\ t\ \le\ \frac{1}{a}\ln\frac{A+D}{c}$. Substituting $a$, $A$, and $D$ explicitly, we find
	\begin{equation}\label{eq:time_bounds_simplified}
		\frac{\sqrt{N}}{\bar\kappa}\,\ln\left[\frac{2}{(2+\bar\kappa)\,c}\right]\ \le\ t\ \le\ \frac{\sqrt{N}}{\bar\kappa}\,\ln\left[\frac{2}{(2-\bar\kappa)\,c}\right].
	\end{equation}
Consequently, the time complexity $\tau$ for the no-click probability to reach a threshold value $c$ satisfies $\mathcal{C}_1 \sqrt{N}\leq \tau \leq \mathcal{C}_2 \sqrt{N}$, with constants $\mathcal{C}_1 \leq \mathcal{C}_2$, establishing the asymptotic time complexity $\tau=\Theta( \sqrt{N})$ \footnote{$\mathcal{C}_1$ and $\mathcal{C}_2$ are not necessarily of order $1$. Their values depend on $c$ and $\bar{\kappa}$. However, the two constants should always be of roughly the same order of magnitude, determined when fixing $c$.}.

\section{Resetting}

\subsection{Resetting at the exceptional point}
Exactly at the exceptional point, the no-click probability behaves like an exponentially decaying function with an oscillation on top whose amplitude decreases with system size. Resetting is neither beneficial nor harmful, as it will give the same time as without resetting. Suppose for large $N$, if $P(t) = e^{-at}$, then after $m$ resets, one has $P_m = P(T)^m = e^{-a mt} = P(mT)$, showing that resetting yields no improvement exactly at the exceptional point.

\begin{figure}[t]
    \centering
    \includegraphics[width=\linewidth]{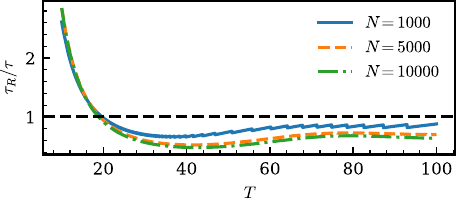}
    \caption{Search time $\tau_R$ (in units of the search time $\tau$ with no resetting) as a function of reset time $T$ for three different values of $N$. Resetting shortens the search time after the curves have crossed the dashed horizontal line. Other parameters are $\bar{r} = 0,\, s = 0.5,\, \bar{\gamma} = 0.9,\, \bar{\kappa} = 1.0, \varepsilon = 0.001$ and stroboscopic interval $\mathrm{d}t = 0.01$.}
    \label{fig:search_time_prefactor}
\end{figure}

However, as mentioned before the exceptional point is highly sensitive to $\bar \gamma$; tuning $\bar{\gamma}\neq 1$ makes the no-click probability sum of more than one exponential. In this regime, one can improve the prefactor by appropriately choosing the reset time $T$, as shown in Fig.~\ref{fig:search_time_prefactor}.

\subsection{Resetting at $T = 1/|\mathrm{Im}\, \lambda_s|$}
\begin{align}
P(T) \approx&\; O_{f} e^{2 \text{Im}\, \lambda_{f} T} + O_{s} e^{2 \text{Im}\, \lambda_{s} T} \\
\approx&\; \frac{1}{N} e^{-2 \frac{|\text{Im}\, \lambda_{f}|}{|\text{Im}\, \lambda_{s}|}} + e^{-2}  \approx e^{-2}\,.
\end{align}
Now, one can further compute $m = \log \epsilon / \log e^{-2} = |\log \epsilon|/2$, which ultimately yields $\tau_R = m T = \frac{1}{|\mathrm{Im} \lambda_s|} |\log \epsilon|/2 = \Theta(1/|\mathrm{Im} \lambda_s|)$. This implies that one cannot benefit from resetting with this particular choice of epoch time $T$.

\subsection{Resetting at $T = 1/|\mathrm{Im}\, \lambda_f|$}
\label{app:reset_fast_mode}

If one reset the system to the initial state $\hat{\varrho}_0$ at every epoch time $T$, and one needs to repeat this process $m$ times to reach a threshold value of $\epsilon$, then the time taken by the reset process is $\tau_R = mT$ and the goal of the problem is to understand the scaling with $N$ and how this compares to the search time $\tau$ of the non-reset process.  By algebraic manipulation, we have $m = \frac{|\log\epsilon|}{|\log P(T)|}$, where 
\begin{eqnarray}
    P(T) \approx O_{f} e^{2 \text{Im}\, \lambda_{f} T} + O_{s} e^{2 \text{Im}\, \lambda_{s} T}\,.
\end{eqnarray}

Note that we ignore the contribution from $O_\pm$, which is negligible if we are not in the exceptional point parameters. Now, let consider the case when we chose $T = \frac{1}{\vert \text{Im}\, \lambda_{f} \vert}$ and use the fact that in leading orders $O_{f} \lesssim 1/N$ and $O_{s} \approx 1$. For this choice,
\begin{eqnarray}
    P(T) &\approx& \frac{e^{-2}}{N} + e^{-2\frac{ |\text{Im} \lambda_{s}|}{|\Im \lambda_{f} |}}\,, \\
    &\approx&  \frac{e^{-2}}{N} +\left(1 - 2 \frac{ |\text{Im} \lambda_{s}| }{ |\text{Im} \lambda_{f} |} \right)\,.
\end{eqnarray}

As we can see in the above equation, the complexity of the resetting procedure depends on which is the dominant contribution after the constant term. In the regime where the third term is dominant, which is the case for regime (D), we will have
\begin{eqnarray}
    m =  \frac{|\log\epsilon|}{|\log P(T)|} \approx  \frac{|\log\epsilon|}{|\log \left( 1 - 2\frac{|\text{Im} \lambda_{s}|}{ |\text{Im} \lambda_{f} |} \right)|} \approx  \frac{|\log\epsilon|}{2 \frac{ |\text{Im} \lambda_{s}|}{  |\text{Im} \lambda_{f}|}+ \ldots }\nonumber\,,\\
\end{eqnarray}
and the corresponding complexity for the resetting process is $\tau_R = mT = |\log \epsilon| \,\, \frac{1}{2|\text{Im} \lambda_{s}|} = \Theta(1/|\mathrm{Im} \lambda_s|)$ which is the same scaling as that for search time-complexity of non-reset process $\tau$.

In the other limit, which occurs in regimes (B) and (C), when we have the term of order $1/N$ is dominant, we will get the scaling
\begin{eqnarray}
    m =  \frac{|\log\epsilon|}{|\log P(T)|} \approx  \frac{|\log\epsilon|}{|\log \left( 1 +\frac{e^{-2}}{N} \right) |}  \approx  \frac{|\log\epsilon|}{\frac{e^{-2}}{N} + \ldots } \,, \nonumber \\
\end{eqnarray}
and the corresponding complexity for the resetting process is $\tau_R = mT = |\log \epsilon| \,\, \frac{N}{|\text{Im}\, \lambda_{f}|}  = \Theta(N/|\mathrm{Im} \lambda_f|)$ which is a different scaling than that for the time-complexity of the non-reset process $\tau$. Since $\text{Im} \lambda_{f} = \Theta(N^{-s})$ for all parameter regimes, the scaling for the reset process will be $\tau_R = \Theta( N^{s+1})$. This can yield better-than-Grover scaling when the monitoring exponent is less than $-1/2$. 

\subsection{Resetting at $ 1/|\Im\, \lambda_f| < T < 1/|\mathrm{Im}\, \lambda_s|$}

 \begin{figure}[!htpb]
    \centering
    \includegraphics[width=\linewidth]{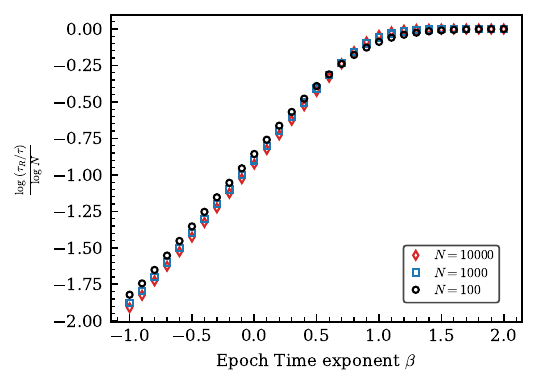}
    \caption{Change in scaling exponent $\alpha_R- \alpha$ plotted as a function of reset time exponent $\beta$.marks parameter values for which resetting improves the scaling exponent ($\alpha_R < \alpha$), while the gray region corresponds to unchanged scaling ($\alpha_R = \alpha$). The parameters are $r = 1.0$, $s = -1.0$, $\epsilon = 0.001$ and $N = 100$ (black circles), $N = 1000$ (blue squares) and $N = 10000$ (red diamonds). This parameter set lies in the hatched region of the phase diagram, where resetting yields an improved scaling.}
    \label{fig:reset_T_intermediate}
\end{figure}

In this section, we present how the time complexity changes in the presence of resetting when the epoch duration $T$ lies between the characteristic times set by the fastest and slowest modes, $1/|\Im \lambda_f| \lesssim T\lesssim 1/|\Im \lambda_s|$. We let the epoch time scale with system size as $T \sim N^\beta$, with $\beta$ interpolating between the scaling exponents associated with the fast and slow modes. Our goal is to determine how resetting modifies the scaling exponent of the search time for intermediate epoch lengths. To this end, we compare the scaling exponent $\alpha_R$ of the reset search time $\tau_R = \Theta(N^{\alpha_R})$ with the exponent $\alpha$ of the non-reset search time $\tau = \Theta(N^\alpha)$. The change in the scaling exponent is obtained from $\frac{\log (\tau_R/\tau)}{\log N}$ which directly yields the quantity $\alpha_R - \alpha$.

 As shown in Fig.~\ref{fig:reset_T_intermediate}, we plotted this difference in scaling exponent $\alpha_R - \alpha$ as a function of $\beta$. Negative values indicate that resetting improves the time complexity by reducing the scaling exponent, while the region where $\alpha_R- \alpha = 0$ corresponds to parameter regimes in which resetting does not alter the asymptotic scaling. Increasing $N$ make the transition sharp at the value of $\beta  = \frac{\log( 1/|\Im \lambda_s|)}{\log N} - 1$. Resetting with the fastest mode $T\sim 1/|\Im \lambda_f|$ offers the largest change in scaling exponent. 

\section{Query complexity} 

To assess the implications of an apparent speedup beyond Grover’s optimality, we analyze the oracle query complexity required to implement Eq.~\eqref{eq:master_equation2} on a unitary, gate-based quantum computer with the current most efficient algorithms simulating Lindbladian systems \cite{Ding_2024}. More generally, the no-fast-forwarding theorem implies that the search time $t$ of the optimal analog protocol lower bounds the query complexity given here by the number $N_{\rm{steps}}$ of the gate-based implementation. In particular, implementing either unitary dynamics~\cite{Berry2007, berry2013gateefficientdiscretesimulationscontinuoustime} or Markovian open-system dynamics~\cite{Childs_2017} for search time $t$ requires $\Omega(t)$ \footnote{$\Omega(\sqrt{N})$  indicates that the complexity can be at least $\sqrt{N}$} queries in the relevant black-box model.

In both the Markovian and non-Markovian cases, better than Grover's speedup in the analog protocol occurs for $s<-0.5$, while in the Markovian case an additional constraint $\bar r \le (2s-1)/4$ must be satisfied. Using the parametrization of the main text, $\kappa=\bar\kappa N^{-s}$ and $\gamma = \bar{\gamma} N^{-\bar{r} - 1}$, we analyze the query complexity of the corresponding unitary simulation. We assume oracle access to the marked element $w$ as $N$ independent i.e.\ $\mathrm{max}(\gamma N ,\kappa )\delta t = \mathcal{O}(1)$. In the algorithm of Ding \textit{et al.}~\cite{Ding_2024}, the timestep scales as $\delta t = \|\mathcal{L}\|^{-1}$, where
\begin{equation}
 \|\mathcal{L}\| = 1 + \|\hat{H}\| + \kappa \|\ket{w}\!\bra{w}\|^2\,, \label{eq:query}
\end{equation}
and $\|\cdot\|$ denotes the operator norm and $H$ following Eq.~\eqref{eq:bare_hamiltonian}. In Table~\ref{tab:query} and Table~\ref{tab:query_reset} we show all asymptotic scalings without and with resetting, respectively. 

\begin{table}[t]
\caption{Query complexity without resetting for $s\leq 0$. Search times $\tau$ are summarized in Table~I of the main text and derived in Sec.~\ref{app:5}. See also Fig.~2(a) of the main text.}
    \centering
    \setlength{\tabcolsep}{4pt}
    \begin{tabular}{c|c|c}
    \hline\hline
      &$s \geq \bar{r}$& $ s \leq \bar{r}$\\
      \hline
      $\delta t \sim ||\mathcal{L}||^{-1}$ & $N^{\bar{r}}$ &$N^{s}$\\
      $\tau = \Theta(N^\alpha)$ & $N^{1+s}$&$N^{2 \bar{r} -s + 1}$\\
      $N_{\rm{steps}} = \frac{\tau}{\delta t}$& $N^{1+s - \bar{r}}$&$N^{2 (\bar{r} -s) + 1}$\\
      $\tau_{\rm{physical}} = N_{\rm{steps}}\sqrt{\delta t}$ &  $N^{1+s - \frac{\bar{r}}{2}} $&$N^{2 (\bar{r} - s) + 1 + \frac{s}{2}}$\\
      \hline
    \end{tabular}
    \label{tab:query}
\end{table}

\begin{table}[t]
    \caption{Query complexity with resetting for $s\leq 0$ and $T = \frac{1}{|\mathrm{Im}\,\lambda_f|}$. The search time in the presence of resetting $\tau_R$ is discussed in Sec.~\ref{app:reset_fast_mode} and is plotted in Fig.~2(b) of the main text.}
    \centering
    \setlength{\tabcolsep}{4pt}
    \begin{tabular}{c|c|c}
    \hline\hline
      &$s \geq \bar{r}$& $ s \leq \bar{r} $ \\
      \hline
      $\delta t  \sim ||\mathcal{L}||^{-1}$ & $N^{\bar{r}}$ &$N^{s}$\\
      $\tau_R = \Theta(N^{\alpha_R})$ & $N^{1+s}$&$N^{1+s}$\\
      $N_{\rm{steps}} = \frac{\tau_R}{\delta t}$& $N^{1+s - \bar{r}}$&$N$ \\
      $\tau_{\rm{physical}} = N_{\rm{steps}}\sqrt{\delta t}$ &  $N^{1+s - \frac{\bar{r}}{2}} $&$N^{1+ \frac{s}{2}}$\\
      \hline
    \end{tabular}
    \label{tab:query_reset}
\end{table}

In the following we assume $s=\bar{r}$, which is the case where the query complexity reaches the minimum. Consequently, the leading contribution to the total number of oracle queries for the search time $\tau = \Theta(N^{\alpha})$ scales as $N_{\mathrm{steps}}=\frac{\tau}{\delta t} \sim N^{\alpha + |s|}=N$. Although the algorithm performs in $N_\mathrm{steps}$, the physical time cost of an algorithmic step is proportional to $\sqrt{\delta t}$~\cite{Ding_2024}. Given any quantum hardware that can operate a single algorithmic step in linear time the physical time necessary to evaluate a single step is thus at least $ \sqrt{\delta t}$ and then the total runtime  $\tau_\mathrm{physical} =N_\mathrm{steps}\sqrt{\delta t} =\Theta( {N^{1-|s|/2}})$. This is minimal when $s=\bar{r}=-1$ and does not violate the optimal bound of Grover. Also, this is consistent with the no fast-forwarding theorem because the query complexity is governed by $N_{\mathrm{steps}} = \tau\,\|\mathcal{L}\|$. In the parameter regimes where the non-unitary dynamics yields a faster physical search time $\tau$, the Lindbladian norm $\|\mathcal{L}\|$
increases with $N$. Consequently, any speedup in $\tau$ is traded for an increased number of oracle calls, rather than achieving it by fast-forwarding the dynamics~\cite{atia_fast-forwarding_2017}.

In the presence of resetting, the dynamics becomes non-Markovian and the search time $\tau_R$ only depends on $s$ in the regime $s \leq 0$. One therefore obtains for case $s=\bar{r}$ the number of steps $N_{\rm{steps}} \sim N^{\alpha_R + |s|}$, which reaches the same conclusion of physical runtime with optimal case $\tau_\mathrm{physical} =\Theta( \sqrt{N})$ being $s=-1$, however, for all considered $\bar{r}$. 

Furthermore, we remark that the simulation time step $\delta t$ vanishes as $N\to\infty$, leading to fundamental simulation challenges, such as the requirement of increasingly high precision with system size and resulting in quantum gate operations with vanishing angle.

 \emph{Data availability.\textemdash}The data that support the findings of this article are openly available at \cite{data_non_hermitian_search}.

\end{document}